\documentclass{emulateapj}
\usepackage{xcolor}
\usepackage[normalem]{ulem}

\mathchardef\mhyphen="2D

\newcommand{\Hb}{\hbox{{\rm H}$\beta$}}
\newcommand{\Mbh}{M_{\rm BH}}

\newcommand{\radlum}{$R_{\rm BLR} - \lambda L_{5100}$}
\newcommand{\Lc}{\hbox{$\lambda L_{5100}$}}
\newcommand{\La}{\hbox{$\lambda L_{5100,\mathrm{AGN}}$}}
\newcommand{\Luv}{\hbox{$\lambda L_{3000}$}}
\newcommand{\less}{\textless}
\newcommand{\more}{\textgreater}
\newcommand{\ergs}{ergs~s$^{-1}$}
\newcommand{\tauoff}{$\tau_{\mathrm{obs}}/\tau_{R-L}$}
\newcommand{\FeII}{\hbox{{\rm Fe}\kern 0.1em{\sc ii}}}
\newcommand{\OIII}{\hbox{{\rm [O}\kern 0.1em{\sc iii}{\rm ]}}}

\newcommand{\red}[1]{{\color{black}{#1}}} 

\begin{document}

\title{The Sloan Digital Sky Survey Reverberation Mapping Project: The H$\beta$ Radius--Luminosity Relation}

\author{Gloria Fonseca Alvarez
 \altaffilmark{1}}
\author{Jonathan R. Trump
 \altaffilmark{1}}
\author{Y. Homayouni
 \altaffilmark{1}}
\author{C. J. Grier
 \altaffilmark{2}}
\author{Yue Shen
 \altaffilmark{3}}
\author{Keith Horne
 \altaffilmark{4}}
\author{Jennifer I-Hsiu Li
 \altaffilmark{3}}
\author{W. N. Brandt
 \altaffilmark{5}
 \altaffilmark{8}
 \altaffilmark{9}}
\author{Luis C. Ho
 \altaffilmark{6}}
\author{B. M. Peterson
 \altaffilmark{7}}
\author{D. P. Schneider
 \altaffilmark{5}}

\altaffiltext{1}{
Department of Physics, University of Connecticut, Storrs, CT 06229, USA
\label{UConn}}

\altaffiltext{2}{
Steward Observatory, The University of Arizona, Tucson, AZ 85721, USA
\label{Arizona}}

\altaffiltext{3}{
Department of Astronomy, University of Illinois at Urbana Champaign, Urbana, IL 61801, USA
\label{Illinois}}

\altaffiltext{4}{
SUPA Physics and Astronomy, University of St. Andrews, Fife, KY16 9SS, Scotland, UK
\label{StAndrews}}

\altaffiltext{5}{
Department of Astronomy \& Astrophysics, The Pennsylvania State University, University Park, PA, 16802, USA
\label{PennState}}

\altaffiltext{6}{
Kavli Institute for Astronomy and Astrophysics, Peking University, Beijing 100871, China
\label{Kavli}}

\altaffiltext{7}{
Department of Astronomy, The Ohio State University, Columbus, OH 43210, USA
\label{OhioState}}

\altaffiltext{8}{
Institute for Gravitation and the Cosmos, The Pennsylvania State University, University Park, PA, 16802, USA
\label{IGC}}

\altaffiltext{9}{Department of Physics, The Pennsylvania State University, University Park, PA, 16802, USA
\label{PennStatePhys}}

%%%%%%%%%%%%%%%%%%%%%%%%%%%%%%%%%%%%%%%%%%%%%%%%%%%%%%%%%%%%%%%%%%%%%%%%%%%%%%%%%%%%%%
% Abstract
%%%%%%%%%%%%%%%%%%%%%%%%%%%%%%%%%%%%%%%%%%%%%%%%%%%%%%%%%%%%%%%%%%%%%%%%%%%%%%%%%%%%%%

\begin{abstract}
Results from a few decades of reverberation mapping (RM) studies have revealed a correlation between the radius of the broad-line emitting region (BLR) and the continuum luminosity of active galactic nuclei. This ``radius--luminosity" relation enables survey-scale black-hole mass estimates across cosmic time, using relatively inexpensive single-epoch spectroscopy, rather than intensive RM time monitoring. However, recent results from newer reverberation mapping campaigns challenge this widely used paradigm, reporting quasar BLR sizes that differ significantly from the previously established radius--luminosity relation. Using simulations of the radius--luminosity relation with the observational parameters of SDSS-RM, we find that this difference is not likely due to observational biases. Instead, it appears that previous RM samples were biased to a subset of quasar properties, and the broader parameter space occupied by the SDSS-RM quasar sample has a genuinely wider range of BLR sizes. We examine the correlation between the deviations from the radius--luminosity relation and several quasar parameters; the most significant correlations indicate that the deviations depend on UV/optical SED and the relative amount of ionizing radiation. Our results indicate that single-epoch black-hole mass estimates that do not account for the diversity of quasars in the radius--luminosity relation could be overestimated by an average of $\sim$ 0.3~dex.

\end{abstract}


%%%%%%%%%%%%%%%%%%%%%%%%%%%%%%%%%%%%%%%%%%%%%%%%%%%%%%%%%%%%%%%%%%%%%%%%%%%%%%%%%%%%%%
% Section 1
%%%%%%%%%%%%%%%%%%%%%%%%%%%%%%%%%%%%%%%%%%%%%%%%%%%%%%%%%%%%%%%%%%%%%%%%%%%%%%%%%%%%%%

\section{Introduction}

Accurate black-hole masses are necessary to understand the growth of black holes and their role in galaxy evolution. In nearby (\less\ 100 Mpc) galaxies, it is possible to measure black-hole mass directly from \red{high spatial resolution observations of} the dynamics of stars and gas \citep[e.g.][]{KormendyHo13}. But for distant active galactic nuclei\footnote{\label{AGN} Throughout this work, we generically use the terms ``quasar'' and ``AGN'' interchangeably to refer to broad-line AGN, as broad lines are necessary for reverberation mapping.} (AGN), the primary method to obtain reliable black-hole masses is reverberation mapping (RM) \red{from time-domain spectroscopy} \citep{Mckee,Peterson04}.

Reverberation mapping measures the time delay between variability in the continuum emission and the corresponding variability in the broad line region (BLR). In the environment around the supermassive black hole, light from the accretion disk is absorbed and re-emitted by the BLR with a delay due to the light travel time between the two emitting regions. The time delay, multiplied by the speed of light, gives a characteristic distance to the BLR,  which is assumed to be in a virial orbit around the black hole. The mass of the black hole is thus given by a virial mass calculation \red{as in Equation (\ref{virial}), using the emission-line broadening ($\Delta\rm V^{2}$), characterized by the line-width FWHM or $\sigma_{\rm line}$, combined with the radius of the BLR}
\begin{equation} \label{virial}
M_{BH} = {f R_{BLR} \Delta V^{2}\over{G}}
\end{equation}

The mass calculation includes a dimensionless factor ``$f$'', to account for the geometry of the orbit and kinematics of the BLR; this factor can be calibrated from comparing RM and dynamical masses \citep{Onken07, Grier13}, the $M_{\mathrm{BH}}-\sigma$ relation \citep{Woo15, Yu19}, or from dynamical modeling of the BLR \citep{Pancoast14}. \red{The f-factor is of order unity and the exact value depends on assumptions like how the broad-line velocity is measured \citep[e.g.,][]{Peterson04,Collin06,Yu19}.}

From RM measurements taken over the last two decades, a correlation has been observed between the measured BLR time delay and the continuum luminosity of the AGN \citep[e.g.,][]{Kaspi00,Bentz09,Bentz13}. From this ``radius--luminosity'' ($R-L$) relation, we can estimate the radius of the BLR with just a luminosity measurement \red{(e.g, Equation \ref{BRL})} and estimate the black-hole mass from single-epoch observations. 
This allows for the measurement of black-hole masses for a large number of AGN without high spatial resolution or long-term monitoring. However, single-epoch estimates are only correct if the $R-L$ relation accurately describes the diverse AGN population; therefore, it is necessary to measure this relation over a broad AGN sample and with the least bias possible.

\cite{Bentz13} used \Hb\ time-lag measurements and reliable subtraction of host galaxy light for 41 AGN from different RM campaigns to determine the following $R-L$ relation between the mean radius of the \Hb-emitting BLR and the AGN continuum luminosity at 5100 \AA\ (\Lc) : 
\begin{equation}\label{BRL}
\log(R_{\rm BLR}/ \mathrm{lt \mhyphen day}) = K + \alpha  \log(\Lc/10^{44}~\mathrm{erg~s^{-1}}) 
\end{equation}
The slope of this relation ($\alpha = 0.533$) is consistent with the $R_{\rm BLR} \propto L^{0.5}$ expectation from basic photoionization models \citep{Davidson72}. \cite{Bentz13} measured an intrinsic scatter in the relation of $\sigma \sim 0.19$, and a normalization $K = 1.527$. The \cite{Bentz13} $R-L$ relation has been the recent standard used to estimate single-epoch black hole masses; however, recent RM results appear to deviate from this relation.

The Sloan Digital Sky Survey Reverberation Mapping (SDSS-RM) project is a dedicated multi-object RM campaign that has been monitoring 849 quasars with spectroscopy and photometry since 2014 \citep{Shen15b}. \cite{Grier17} published an \Hb\ $R-L$ relation for 44 AGN from the first year of SDSS-RM monitoring. The time lags measured by SDSS-RM are often significantly shorter than those predicted by Equation (\ref{BRL}) for their given AGN luminosity, and thus these sources fall below the \cite{Bentz13} $R-L$ relation. In addition, the Super-Eddington Accreting Massive Black Holes (SEAMBH) survey presented a $R-L$ relation for a sample of rapidly-accreting AGN that also differs from \cite{Bentz13} in the same manner \citep{Du16,Du18,Du19}.

In this work we examine if this discrepancy is due to observational biases that restrict the allowable lag detections, or if the SDSS-RM and SEAMBH samples have properties that \red{represent a broader population of AGN} compared to previous RM studies; thus indicating a physical origin for the discrepancy\red{, as suggested by recent work \citep{Czerny19,Du19}}. We explore this by simulating a $R-L$ relation based on \cite{Bentz13}, while imposing the observational constraints of the SDSS-RM dataset. We present the data included in our study in Section 2, and provide a detailed description of our simulated $R-L$ relation and results in Section 3. In Section 4, we discuss possible causes for the discrepancy. Throughout this work we assume a standard $\Lambda$CDM cosmology with $\Omega_\Lambda = 0.7$, $\Omega_M = 0.3$, and $H_0 = 70\ \mathrm{km~s^{-1}~Mpc^{-1}}$.

%%%%%%%%%%%%%%%%%%%%%%%%%%%%%%%%%%%%%%%%%%%%%%%%%%%%%%%%%%%%%%%%%%%%%%%%%%%%%%%%%%%%%%
% Section 2
%%%%%%%%%%%%%%%%%%%%%%%%%%%%%%%%%%%%%%%%%%%%%%%%%%%%%%%%%%%%%%%%%%%%%%%%%%%%%%%%%%%%%%

\section{Data}

For our analysis, we compare \Hb\ lags, \Lc, and the best-fit \radlum\ relation for the \cite{Bentz13}, \cite{Grier17}, and \cite{Du16,Du18} datasets. The lags for the 3 RM campaigns were measured using different methods: \cite{Bentz13} and \cite{Du16,Du18} used the interpolated cross-correlation function (ICCF, \citealp{Gaskell87, White94, Peterson04}), while \cite{Grier17} primarily used JAVELIN \citep{Zu11} and CREAM \citep{Starkey16}. JAVELIN and CREAM use different assumptions than ICCF but are designed to produce similar results, so any deviations from the \cite{Bentz13} $R-L$ relation should not be due to the different lag detection methods, \red{as discussed in section 2.3}. We briefly describe the details of the lag measurement methods in section 2.1. 

%%%%%%%%%%%%%%%%%%

\begin{figure}[!t]
\centering
\includegraphics[width=\columnwidth]{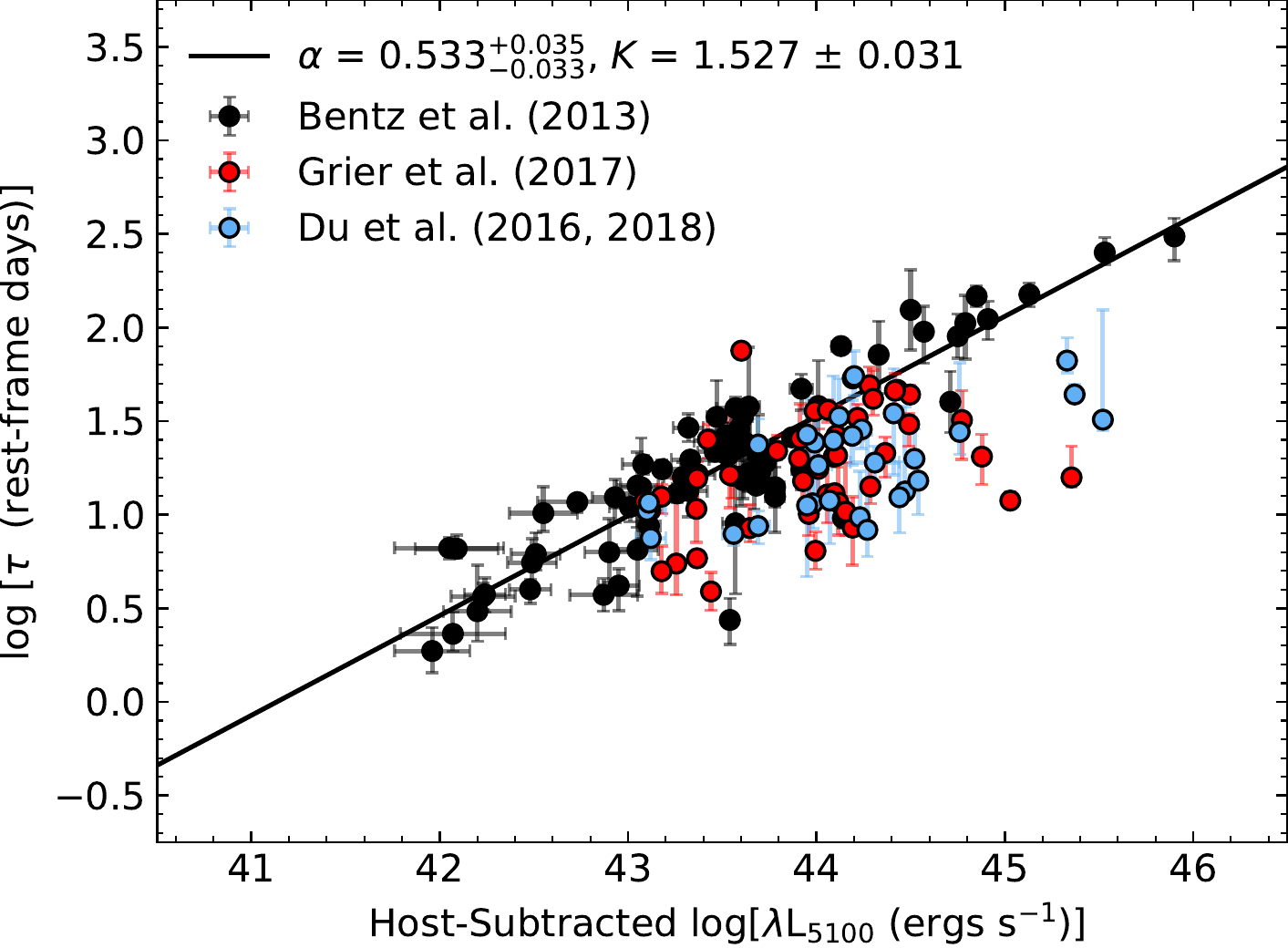}
\caption{The $R-L$ relation for \Hb\ time lags from \cite{Bentz13}, \cite{Grier17}, and \cite{Du16, Du18}. The black line shows the $R-L$ relation from \cite{Bentz13}, with a slope $\alpha = 0.533$ and a normalization $K = 1.527$. The lag measurements from SDSS-RM \citep{Grier17} and SEAMBH \citep{Du18} frequently lie below the $R-L$ relation established by \cite{Bentz13}.}
\label{bentzfit}
\end{figure}

%%%%%%%%%%%%%%%%%%

\begin{figure}[!t]
\centering
\includegraphics[width=\columnwidth]{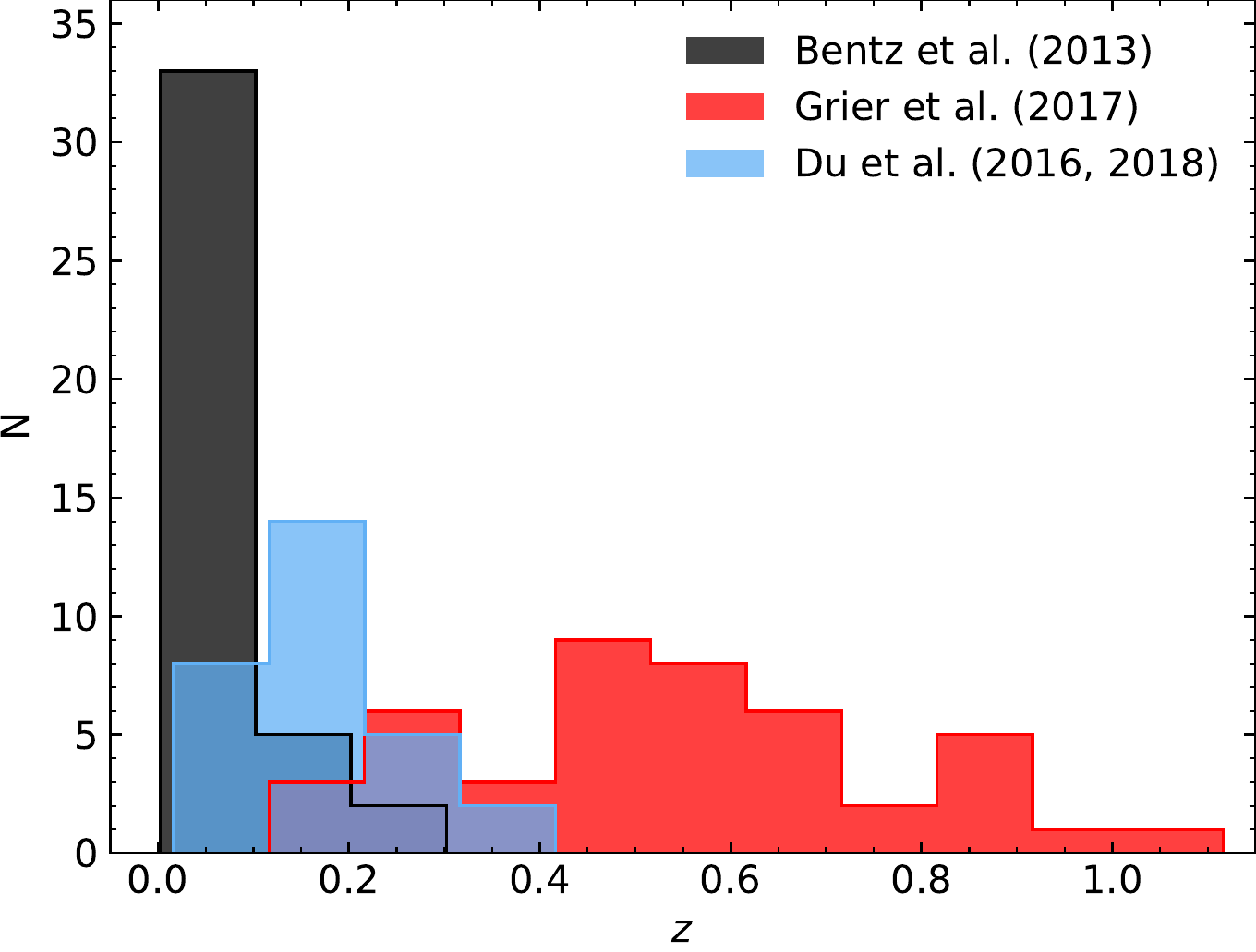}
\includegraphics[width=\columnwidth]{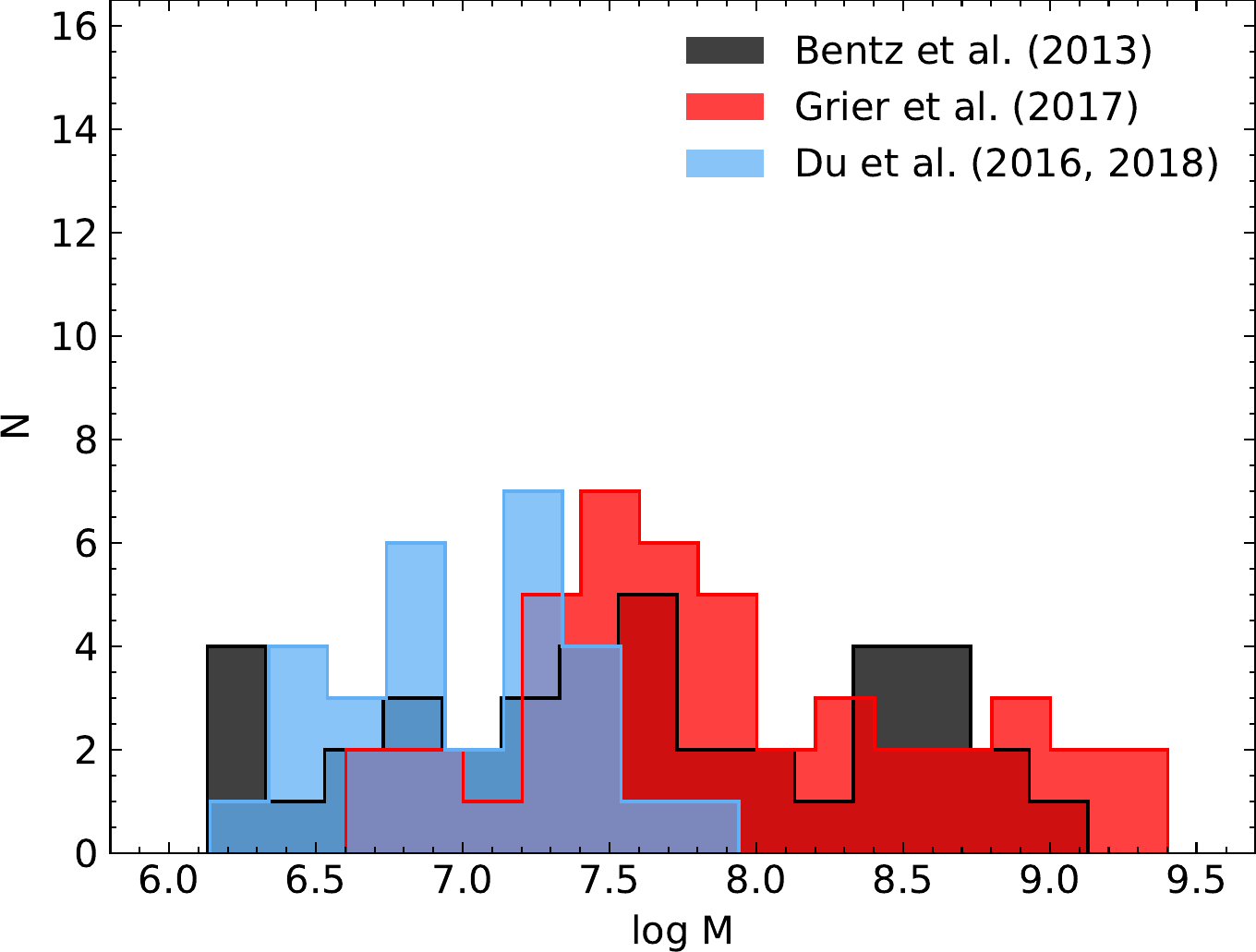}
\includegraphics[width=\columnwidth]{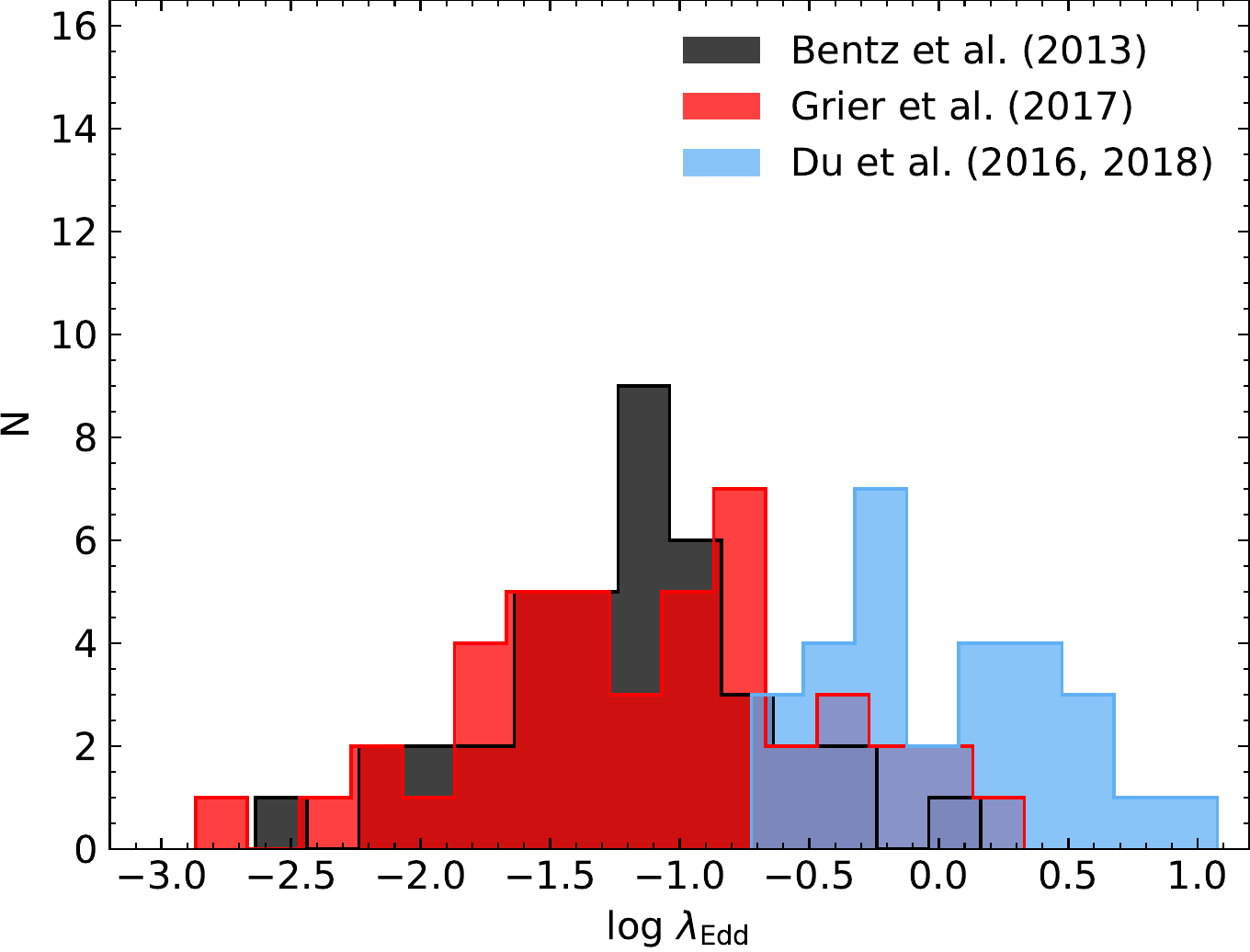}
\caption{\red{From top to bottom: Distribution of redshift, mass, and $\lambda_{\rm Edd}$  for the \cite{Bentz13}, \cite{Grier17} and \cite{Du16,Du18} samples.} }
\label{SampleHisto}
\end{figure}

Figure \ref{bentzfit} presents the $R-L$ relation for the \cite{Bentz13}, \cite{Grier17}, and \cite{Du16,Du18} samples of AGN with \Hb\ RM lags. We describe these three samples in detail in the subsections below. \red{The distribution of AGN properties in each sample is presented in Figure \ref{SampleHisto}. For the Eddington ratio ($\lambda_{\rm Edd} = {L_{\rm bol}\over{L_{\rm Edd}}}$), we assume $L_{\rm bol}$ = 5.15\Luv\ and $L_{\rm bol} = 9.26\lambda L_{5100}$ \citep{Richards06}. Published 3000 \AA\ luminosities are available only for 41 of the \cite{Grier17} AGN; we use the 5100 \AA\ luminosities for all other AGN in the three samples. We use black-hole masses for the \citet{Bentz13} sample from the compilation of \citet{Bentz15}.} In all three samples, the AGN luminosities are host-subtracted, and as such the luminosity uncertainties include a contribution from the uncertainty associated with the host-galaxy decomposition. In general this means that the AGN luminosity uncertainties are largest for low-luminosity and host-dominated AGN, and are generally small for luminous AGN. \red{We determine the best-fit $R-L$ relation for each sample employing multiple linear regression with the python MCMC software \texttt{PyMC3}, including uncertainties in both radius (y-axis) and luminosity (x-axis) and allowing for excess intrinsic scatter.}

\subsection{Lag Measurement Methods}

The ICCF determines the cross-correlation between two light curves, measured as the Pearson correlation coefficient $r$ as a function of time delay $\tau$. Because the data are unevenly spaced due to observational constraints, the ICCF linearly interpolates the first light curve to produce overlapping points to calculate $r$ for any delay $\tau$. The same process is repeated starting with the second light curve shifted by $-\tau$. The cross-correlation coefficient for a given $\tau$ is obtained by averaging the two values of $r$. The ICCF repeats this procedure for a range of $\tau$, to obtain the final cross-correlation function (CCF). The likely time lag between the two light curves is given by the centroid of the CCF. The uncertainties are calculated using Monte Carlo methods with flux re-sampling and random subset sampling \citep{Peterson04}.

%%%%%%%%%%%%%%%%%%

%%%%%%%%%%%%%%%%%

Instead of using linear interpolation, JAVELIN assumes that the variability of the continuum light curve is best described by a damped random walk (DRW) model. JAVELIN then models the BLR light-curve response with the same DRW model combined with a top-hat transfer function centered at a lag $\tau$, producing a BLR light curve model that is a shifted, smoothed, scaled version of the continuum light curve. Markov Chain Monte Carlo (MCMC) is used to identify the most likely lag and uncertainty. CREAM adopts a similar approach to JAVELIN to measure lags, with the same DRW assumption about variability, but with a slightly different treatment of the uncertainties. \red{Detailed simulations by \cite{Li19} and \cite{Yu20} find that, for lightcurves of similar cadence and noise to SDSS-RM, JAVELIN produces more accurate lags and lag uncertainties than ICCF, and fewer false positives.} \cite{Grier17} measured \Hb\ lags using JAVELIN, ICCF, and CREAM; in this work we primarily utilize the lags from JAVELIN and CREAM, while noting that the ICCF lags of SDSS-RM quasars produce the same \red{offset in the} $R-L$ relation (Figure \ref{sdssiccf}). 

\subsection{Bentz et al.}

\cite{Bentz13} collected a sample of 41 AGN from previous RM surveys, focusing on adding accurate host-galaxy subtraction from HST imaging. The sample primarily includes nearby AGN that were generally selected to be apparently bright and variable, with luminosities in the range $10^{42} < \La < 10^{46}$\ \ergs. The AGN have lags measured from observing campaigns with monitoring durations that ranged from 64 to 120 days, with cadences as rapid as 1 day between observations. Lags were measured using the ICCF method, resulting in 70 \Hb\ time lags for 41 unique AGN in the range 2--100 rest-frame days.

The luminosity measurements are corrected for host-galaxy contributions; this is especially important for lower-luminosity AGN since galaxy contamination leads to an overestimation of \Lc, steepening the $R-L$ relation. Previous RM surveys that did not correct for host-galaxy luminosity found a steeper $R-L$ relation with a slope $\alpha \sim$ 0.70 \citep{Kaspi00}. \cite{Bentz13} measured the host-galaxy contribution for each AGN through morphological decomposition of HST/ACS images, using the GALFIT software \citep{Peng02} to determine the best-fit point-source AGN and extended galaxy surface brightness profiles implementing a nonlinear least-squares fit algorithm.

Figure 11 in \cite{Bentz13} presents the $R-L$ relation observed for their measured \Hb\ time lags, with a slope $\alpha = 0.533^{+0.035}_{-0.033}$ and a normalization $K = 1.527^{+0.031}_{-0.031}$ for the best-fit line. Our fitting method yields a nearly identical slope \red{$\alpha = 0.52 \pm 0.03$ and a normalization $K = 1.52 \pm 0.03$} for the \cite{Bentz13} \Hb\ lags.

\subsection{SDSS-RM}

\cite{Grier17} successfully measured \Hb\ time lags for 44 AGN from the SDSS-RM survey. The AGN have luminosities $10^{43} < \La < 10^{45.5}$~\ergs\ and redshifts 0.12 \less\ {\it{z}} \less\ 1. The full SDSS-RM sample is magnitude-limited (by $i_{\rm AB}<21.7$), with no other selection criteria for AGN properties. This results in a sample that is more representative of the general AGN population, and a greater diversity in redshift and other AGN properties compared to previous RM studies. For example, the SDSS-RM sample spans a much broader range of emission-line widths, strengths, and blueshifts compared to the sample of \cite{Bentz13} \citep[see Figure 1 of][]{Shen15b}.

Spectra of the quasars were obtained using the Baryon Oscillation Spectroscopic Survey (BOSS) spectrograph \citep{Smee13} on the SDSS 2.5 m telescope \citep{Gunn06} at Apache Point Observatory. The initial observations include 32-epochs taken over a period of 6 months in 2014. The exposure time for each observation was $\sim$ 2 hr and the average time between observations was 4 days (maximum 16.6 days). 

Photometric observations were acquired in the {\it{g}} and {\it{i}} filters with the Bok 2.3 m telescope and the Canada-France-Hawaii Telescope (CFHT). Additionally, synthetic photometric light curves were produced from the BOSS spectra in the {\it{g}} and {\it{i}} bands. All of the {\it{g}} and {\it{i}} band light curves were merged using the CREAM software \citep{Starkey16} to create a continuum light curve for each AGN \citep[see][for additional details of the light-curve merging procedure]{Grier17}.

%%%%%%%%%%%%%%%%%%

\begin{figure}[!t]
\centering
\includegraphics[width=\columnwidth]{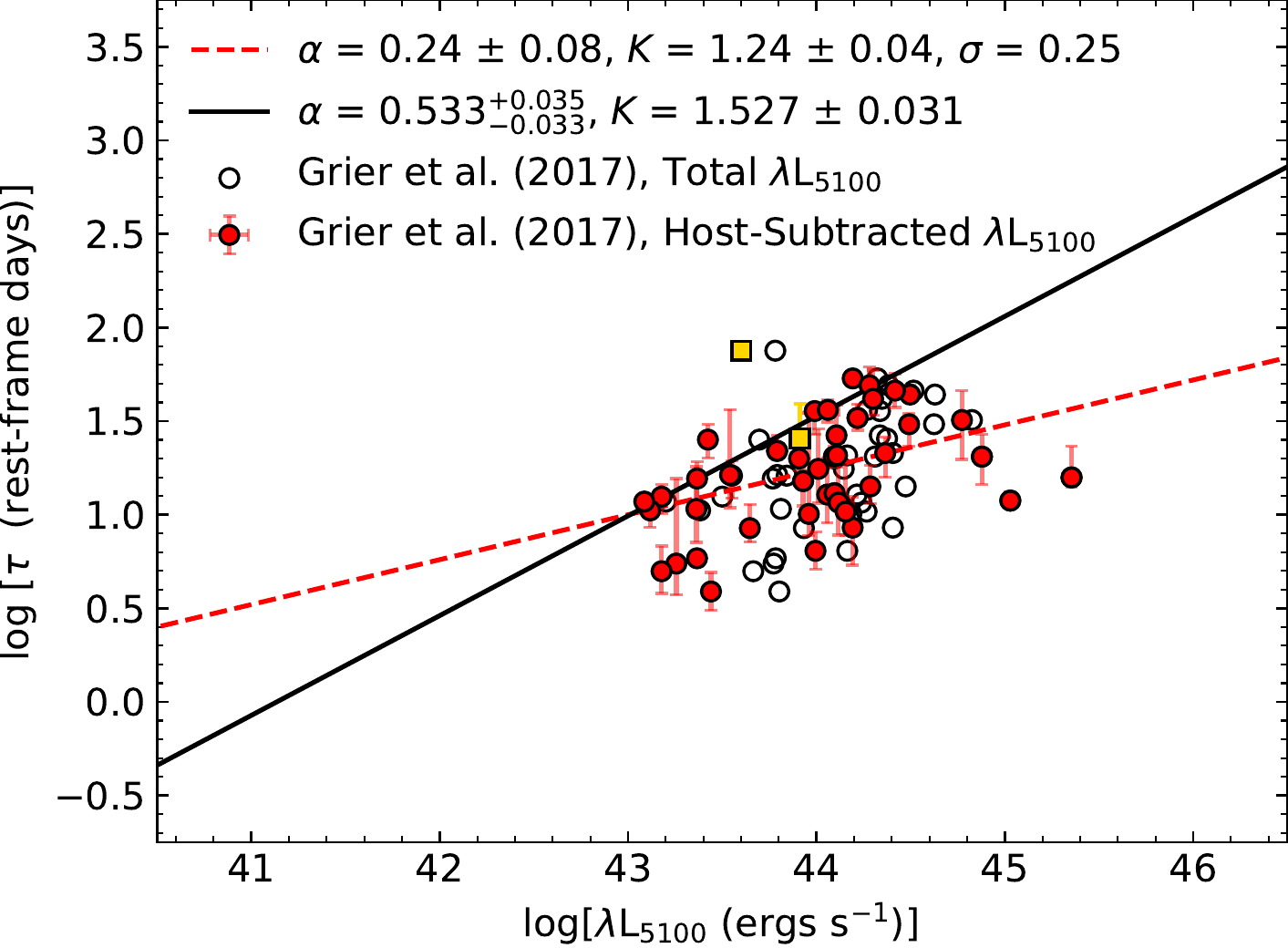}
\caption{The $R-L$ relation for 44 AGN in the SDSS-RM survey, with \Hb\ time lags from \cite{Grier17} and \Lc\ from \cite{Shen15a}. Out of the 44 lags, 32 were measured using JAVELIN and 12 were measured using CREAM. The open circles have \Lc\ that includes host-galaxy light, while the solid red circles have AGN luminosities (\Lc) that are host-subtracted using principal component analysis of the coadded spectra. Our best-fit line for the red (host-subtracted) points is shown as the red dashed line, with a slope $\alpha$ = 0.24 $\pm$ 0.08 and a normalization of $K=1.24 \pm 0.04$ that both differ from the \cite{Bentz13} best-fit $R-L$ relation (shown as the black solid line) by $>$ 3$\sigma$. The two \red{square} points were excluded from the fitting (see text for details). The SDSS-RM AGN generally have lags that are shorter than expected from the \citet{Bentz13} $R-L$ relation at a given host-subtracted $\Lc$.}
\label{sdssfit}
\end{figure}

%%%%%%%%%%%%%%%%%%

\cite{Grier17} measured \Hb\ reverberation lags using ICCF, JAVELIN and CREAM. Each method used a lag search range between $-100$ and 100 days, given the length of the SDSS-RM observation baseline ($\sim$ 200 days). This resulted in 32 lags from JAVELIN and 12 from CREAM, only including ``reliable'' positive time lags that have ${\rm SNR} > 2$, a single well-defined peak in the lag probability distribution function, and a correlation coefficient of $r_{\rm max}>0.45$.

\cite{Shen15b} used principal component analysis (PCA) to decompose the quasar and host-galaxy spectra, assuming that the total spectrum is a combination of linearly independent sets of quasar-only and galaxy-only eigenspectra. The SDSS eigenspectra are taken from \cite{Yip04}. To obtain the quasar-only spectrum, \cite{Shen15b} subtracted the best-fit host-galaxy spectrum from the total spectrum. \cite{Yue18} independently estimated the host-galaxy contribution using imaging decomposition and found consistent results to the spectral decomposition.

Figure \ref{sdssfit} presents the relation between the 44 SDSS-RM \Hb\ time lags and \Lc. Host-subtracted continuum luminosity (\Lc) measurements were taken from \cite{Shen15a}. The points in red represent AGN luminosities that are host-subtracted as described above. The observed rest-frame time lags are generally shorter than predicted from the \cite{Bentz13} $R-L$ relation. The SDSS-RM data exhibit a positive correlation between radius and luminosity, with a Spearman's $\rho=0.54$ and a null probability of no correlation of $p\sim 0.0$. The $R-L$ properties of the SDSS-RM quasars are best fit by a line with shallower slope \red{and lower normalization}, as shown as the red best-fit line of slope \red{$\alpha = 0.24 \pm 0.08$ and a normalization $K = 1.24 \pm 0.04$.} However, the limited dynamic range of the SDSS-RM quasars means that the data could also be consistent with the same $\alpha \simeq 0.5$ slope of the \citet{Bentz13} data, with an average offset of shorter lags in SDSS-RM quasars over a range of continuum luminosities. Fitting the same SDSS-RM data, while fixing the slope to be 0.533, results in the same \red{lower normalization K = 1.24 $\pm$ 0.05.} For this and all subsequent least-squares fitting, we exclude the SDSS-RM data point with the longest lag and smallest fractional uncertainty as an outlier (RMID 781). We also exclude the hyper-variable quasar RMID 017, as it increases in luminosity by a factor of $\sim$10 over the span of the SDSS-RM monitoring \citep{Dexter19}.

%%%%%%%%%%%%%%%%%%

\begin{figure}[!t]
\centering
\includegraphics[width=\columnwidth]{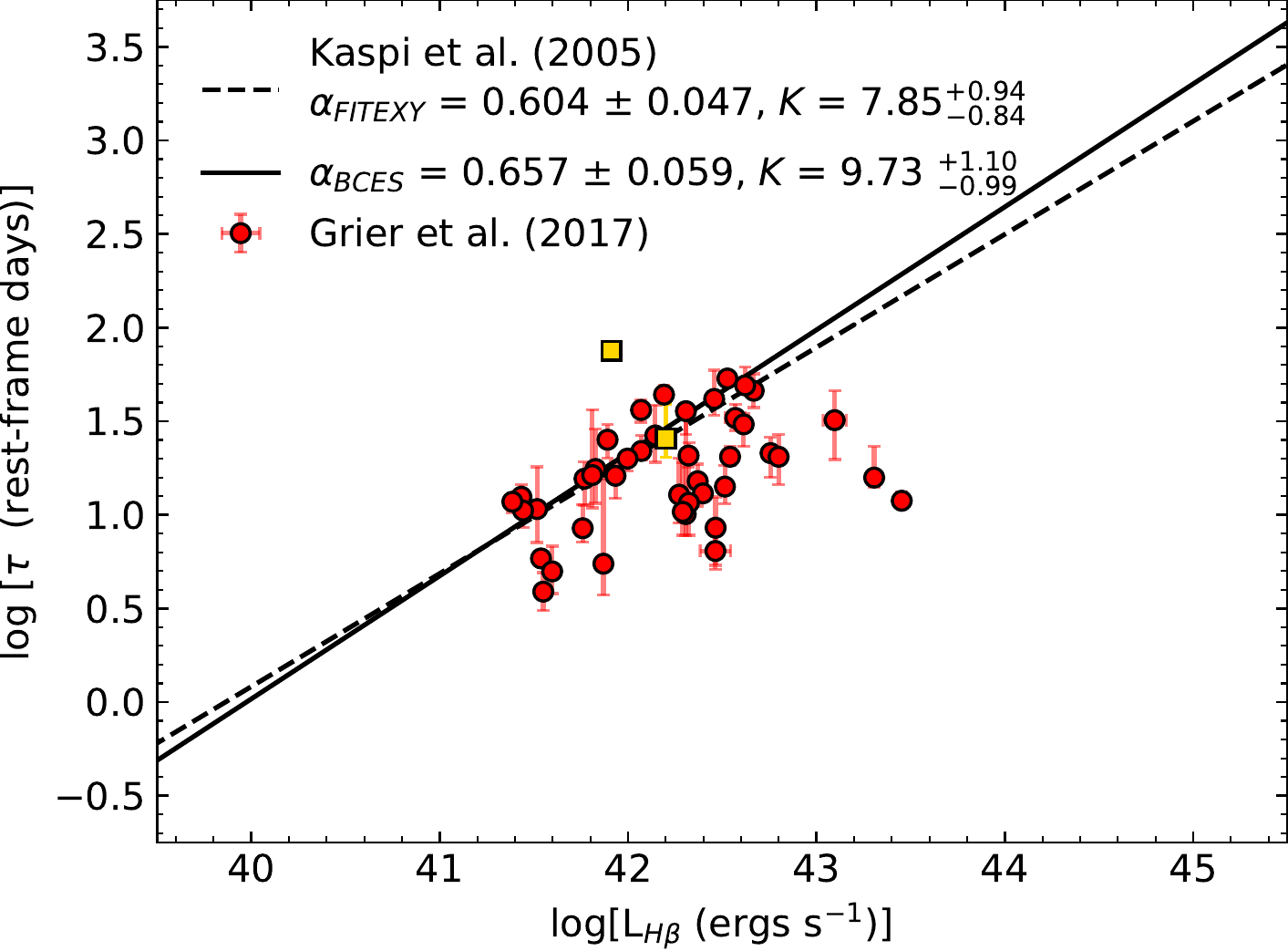}
\caption{The $R-L$($\Hb$) relation for the 44 SDSS-RM AGN, with \Hb\ time lags from \cite{Grier17} and broad-line $\Hb$ luminosity from \cite{Shen19}. Black solid and dashed lines show the relation between \Hb\ time lags and  L$_{\rm H\beta}$ from \cite{Kaspi05} for two different fitting methods. The SDSS-RM AGN have lags that fall below the $R-L$(\Hb) relation.}
\label{sdsshbeta}
\end{figure}

%%%%%%%%%%%%%%%%%%

Figure \ref{sdssfit} also includes the total \Lc\ without host-galaxy subtraction for each AGN as open circles, as an indication of the typical relative contribution of AGN and galaxy light.
We further demonstrate that the $R-L$ offset is not due to under-subtracted host-galaxy luminosities by examining the $R-L$(\Hb) relation, presented in Figure \ref{sdsshbeta}. The luminosity from the \Hb\ emission line is produced by the AGN broad-line region and does not have any galaxy contribution. Since the \cite{Bentz13} sample lacks published \Hb\ luminosities, we cannot compare that sample with the SDSS-RM $R-L$(\Hb) relation. Instead, we use the \cite{Kaspi05} best-fit $R-L$(\Hb) lines that were fit to a subset of the \cite{Bentz13} data, shown as dashed and solid lines in Figure \ref{sdsshbeta}. The SDSS-RM lags show the same general trend of falling below the relation measured from previous RM data.

\begin{figure}[!t]
\centering
\includegraphics[width=\columnwidth]{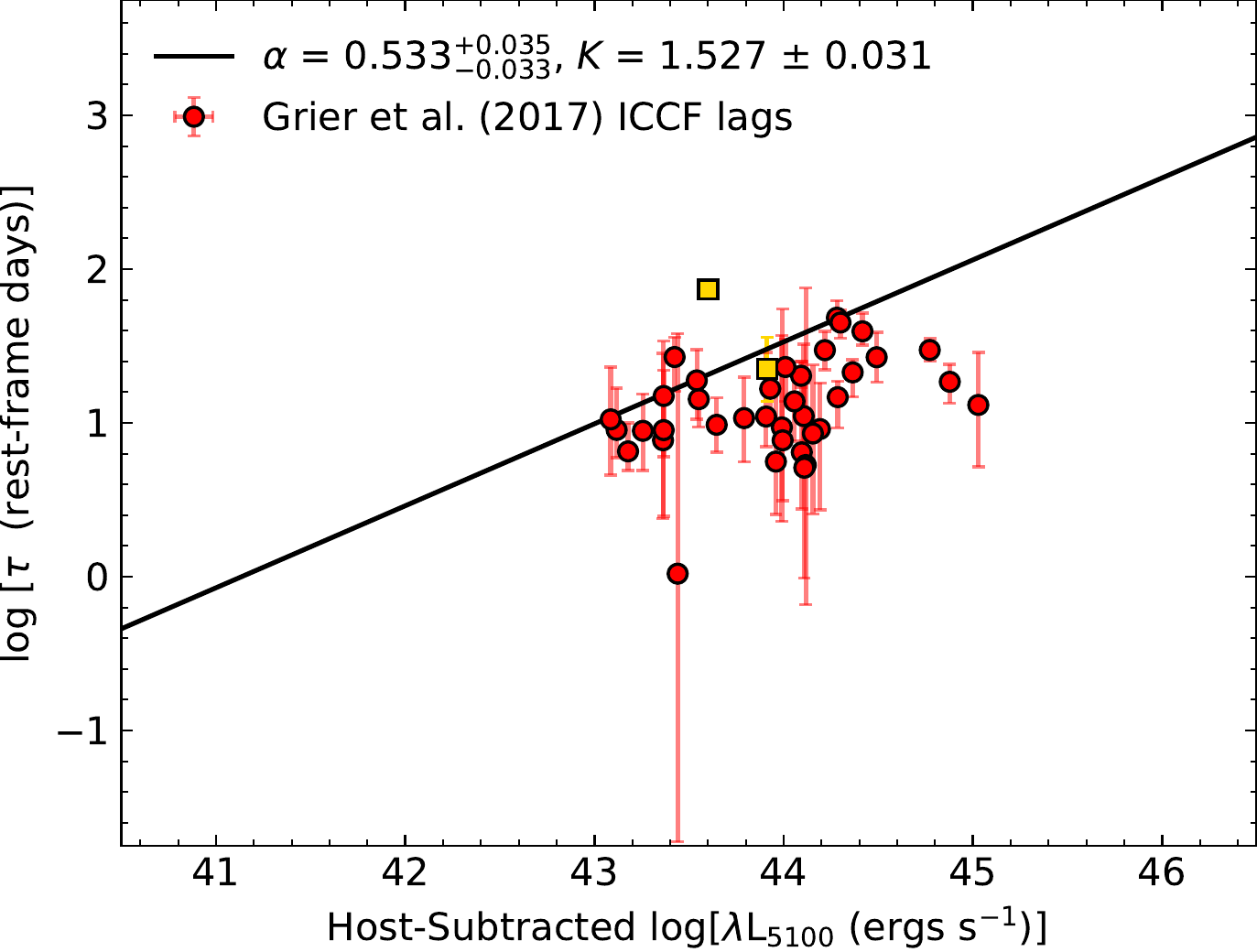}
\caption{The $R-L$ relation for 39 ICCF lags of the SDSS-RM AGN from \cite{Grier17} with host-subtracted \Lc\ from \cite{Shen15a}. Five AGN have ICCF lags less than 1 day and are not shown in the figure. The ICCF lags of SDSS-RM AGN have the same offset from the \cite{Bentz13} $R-L$ relation seen in Figure \ref{sdssfit}.}
\label{sdssiccf}
\end{figure}

Finally, to be certain that the different lag-detection methods are not the cause of the offset, we present the $R-L$ relation using ICCF measured lags from SDSS-RM in Figure \ref{sdssiccf}. The ICCF lags fall below the \cite{Bentz13} relation just as seen in the JAVELIN and CREAM lags. 

%%%%%%%%%%%%%%%%%%

%%%%%%%%%%%%%%%%%%

\subsection{SEAMBH}

The SEAMBH project is a RM campaign spanning 5 years of monitoring \citep{Du16,Du18}. The AGN in the sample were selected from SDSS using a dimensionless accretion rate $\dot{\mathcal{M}}$, derived from the standard thin-disk equations \citep{Wang14b}:
\begin{equation}\label{mscript}
\dot{\mathcal{M}} = 20.1 {\left( L_{44}\over{\cos i} \right)^{3/2}} m_{7}^{-2}
\end{equation}

The inclination of the disk is given by $i$ and we assume cos $i$ = 0.75 \citep{Du18}.
\red{The luminosity and mass dependence are parameterized as L$_{44} = \rm \Lc / 10^{44} $ and m$_7 = \rm M /{10^{7}\ M_{\odot}}$, respectively}.
The SEAMBH AGN were selected to have $\dot{\mathcal{M}} > 3$; the sample of 29 AGN has $10< \dot{\mathcal{M}} <10^{3}$, giving them higher accretion rates than the general AGN population. \red{For comparison, the \cite{Bentz13} and \cite{Grier17} samples have a median  $\dot{\mathcal{M}}$ \less\ 0.50.} Spectroscopic and photometric observations were made over 5 years with the Lijiang 2.4 m telescope, averaging 90 nights per object. Typical exposure times were 10 minutes for photometry and 1 h for spectroscopy. \cite{Du16,Du18} used an empirical relation to determine the host-galaxy contribution to the spectrum based on \Lc, derived by \cite{Shen11} for SDSS fiber spectra:
\begin{equation}\label{Lest}
{L_{5100}^{\rm host} \over L_{5100}^{\rm AGN}} = 0.8052 - 1.5502x + 0.912x^2 - 0.1577x^3
\end{equation}
Here $x = L^{\mathrm{tot}}_{5100} \times \ 10^{-44}$~\ergs. For spectra with $L^{\mathrm{tot}}_{5100} > 1.053 \times 10^{44}$~\ergs, the host-galaxy contribution was assumed to be zero.

\begin{figure}[!t]
\centering
\includegraphics[width=\columnwidth]{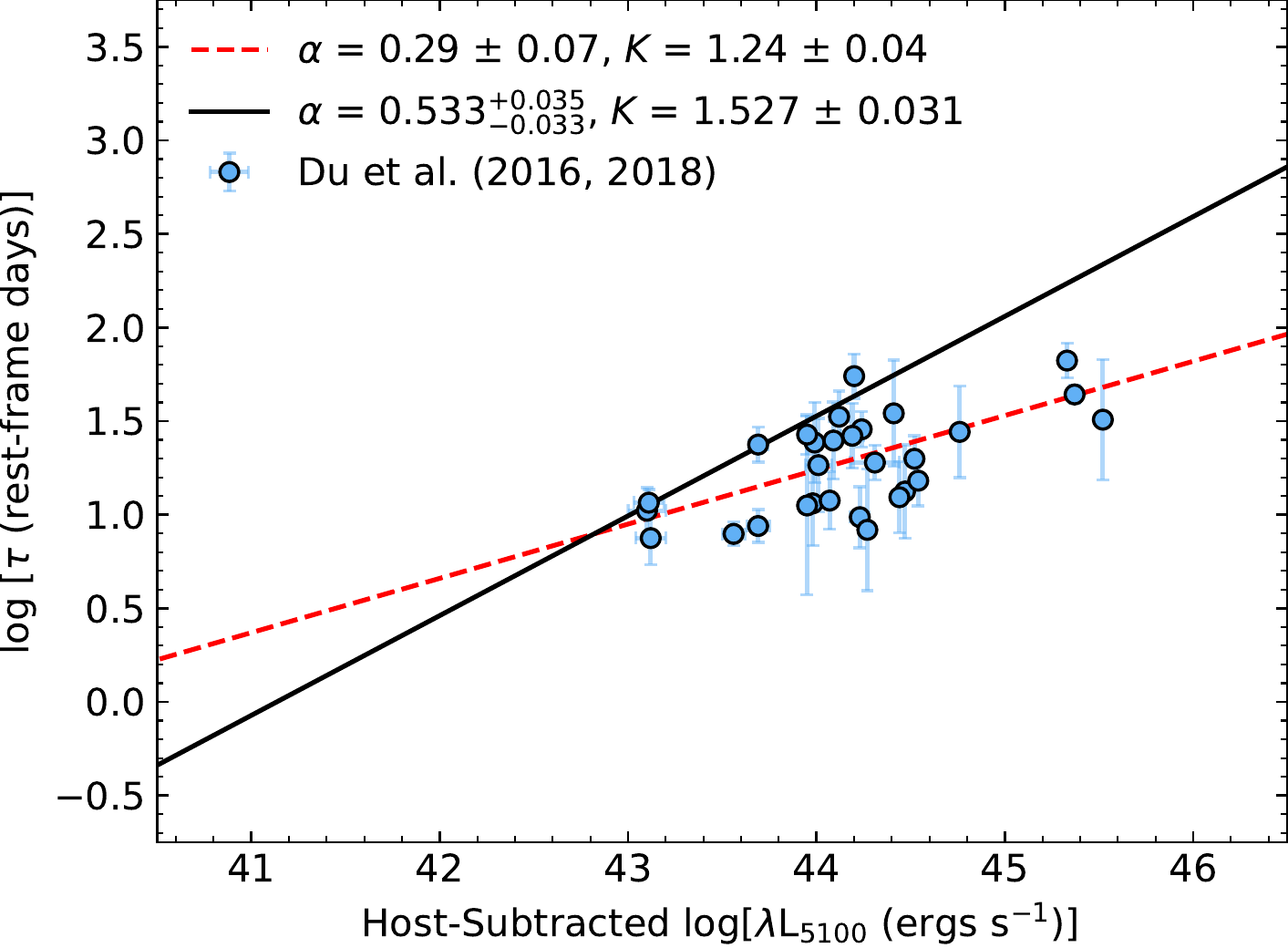}
\caption{The $R-L$ relation for the 29 \Hb\ time lags measured by \cite{Du16,Du18}. The time lags were measured using ICCF and include 19 lags from \cite{Du16} and 10 lags from \cite{Du18}. The AGN luminosities (\Lc) were calculated using a galaxy-contribution estimate based on Equation (\ref{Lest}). Our best-fit line, shown as a dashed red line, gives a slope $\alpha = 0.29 \pm 0.07$ and a normalization $K = 1.24 \pm 0.04$, indicating that the SEAMBH AGN (like the SDSS-RM AGN) follow a relation that is significantly below the previous \cite{Bentz13} $R-L$ relation.}
\label{dufit}
\end{figure}

The $R-L$ relation for the 29 SEAMBH \Hb\ lags measured by \cite{Du16,Du18} is presented in Figure \ref{dufit}. Similar to the SDSS-RM data in Figure \ref{sdssfit}, the measured lags are shorter than expected from Equation (\ref{BRL}), resulting $R-L$ relation with a shallower slope \red{$\alpha = 0.29 \pm 0.07$ and a lower normalization $K = 1.24 \pm 0.04$}. The SEAMBH data, like the SDSS-RM data, cover a limited dynamic range on both axes, and also appear consistent with a slope of $\alpha \simeq 0.5$ with an average offset for shorter lags over a broad range of continuum luminosity.

%%%%%%%%%%%%%%%%%%

%%%%%%%%%%%%%%%%%%%%

%%%%%%%%%%%%%%%%%%%%%%%%%%%%%%%%%%%%%%%%%%%%%%%%%%%%%%%%%%%%%%%%%%%%%%%%%%%%%%%%%%%%%%
% Section 3
%%%%%%%%%%%%%%%%%%%%%%%%%%%%%%%%%%%%%%%%%%%%%%%%%%%%%%%%%%%%%%%%%%%%%%%%%%%%%%%%%%%%%%

\section{Simulating Observational Bias on the $R-L$ relation}

\red{The effects of the SDSS-RM observational limits on the observed $R-L$ relation are not easily predictable. For instance, the sample is magnitude-limited in the $i$ band, rather than limited by the luminosity used for the R-L relation. There are also constraints on the length of the measurable lags due to the duration and cadence of the observations.} In order to examine how observational biases affect the $R-L$ relation, we simulated a $R-L$ relation starting from \cite{Bentz13} \red{(Equation \ref{BRL})} and including observational errors and limits appropriate for the SDSS-RM monitoring campaign.

\subsection{General Simulation} 

To create a representative sample of AGN, we generated $10^7$ random AGN luminosities in the range 10$^{42}$--10$^{46}$~\ergs\ following the \red{$i$-band} luminosity function from \red{\cite{Ross13}}:
%($\alpha = 3.37, \beta = 1.16$)
\begin{equation}\label{LD}
\Phi = {\Phi^*\over{ \left(L/ L_{B}^*\right)^{3.37}   + \left(L/L_B^* \right)^{\red{1.16}} } }
\end{equation}
The $L^{3.37}$ and \red{$L^{1.16}$} terms represent the bright and faint end of the distribution, respectively, with a break luminosity \red{$L_{B}^* = 10^{44.62}$}~\ergs. %\blue{Describe why this is a good LF for this sample. } 
\red{This results in a distribution of AGN luminosities in the observed $i$-band. To shift the observed $i$-band luminosities to \Lc\, we use the average quasar SED of \cite{Richards06} and a randomly assigned redshift. Each simulated AGN was assigned a redshift randomly drawn from the set of 44 SDSS-RM AGN and spanning $0.2<z<1.2$.}

%%%%%%%%%%%%%%%%%%

\begin{figure}[!t]
\centering
\includegraphics[width=\columnwidth]{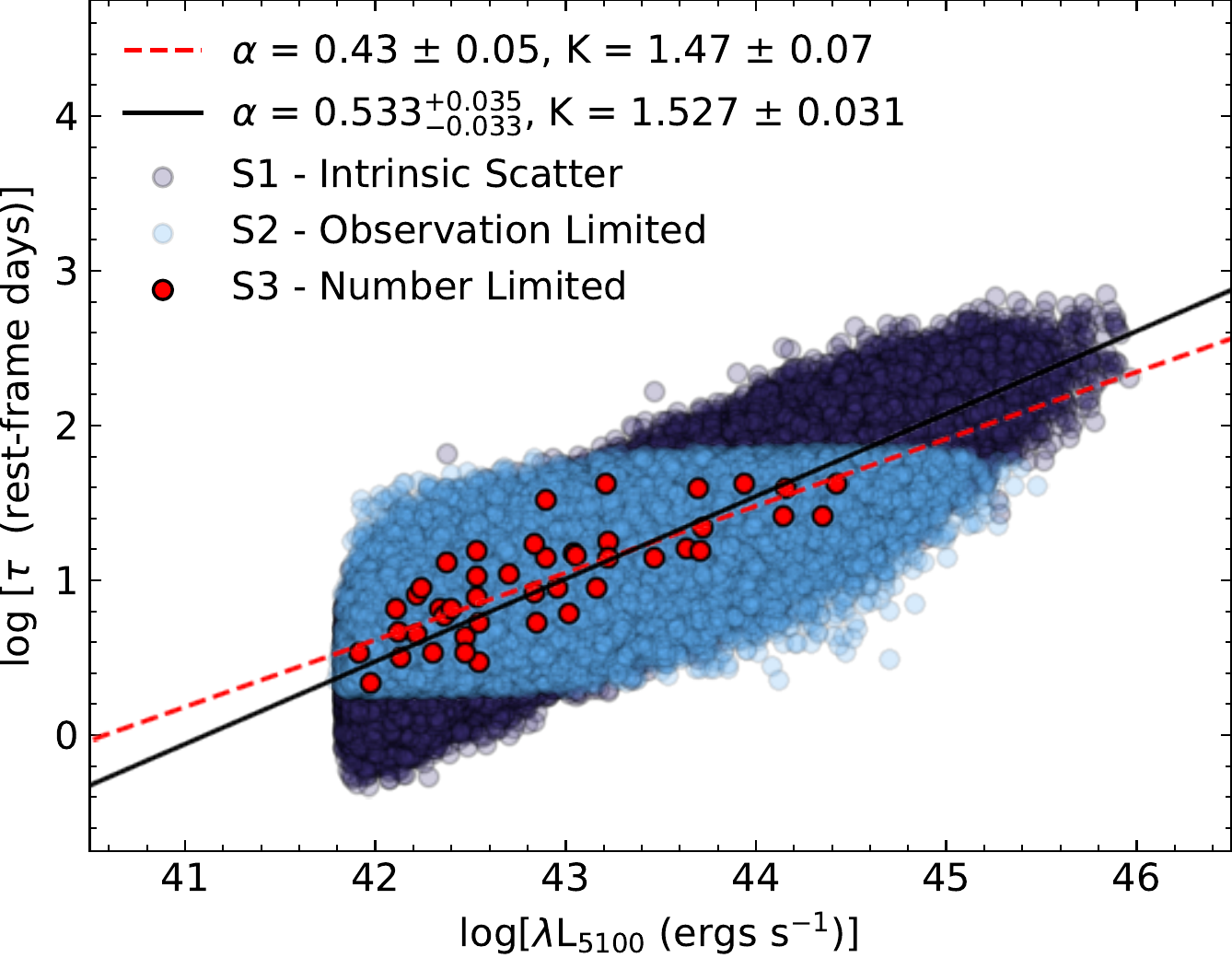}
\caption{One iteration of the simulated \Hb\ $R-L$ relation. Points in purple represent the relation for AGN in sample S1, which includes only the intrinsic scatter in \cite{Bentz13}. Points in blue represent the AGN in sample S2, which takes into account observational errors and observational limits typical of SDSS-RM. The points in red are 44 random points chosen from sample S2, this accounts for the number of lags detected by SDSS-RM. The red line shows the best fit for the points in red (S3).}
\label{simulation}
\end{figure}

%%%%%%%%%%%%%%%%%%

We calculated the expected radius of the \Hb\ BLR (given as $\tau = R/c$ in days) for each \red{\Lc} using the \cite{Bentz13} relation, including an intrinsic scatter of $\sigma_{\rm int} = 0.19$. The BLR radius for each of the $10^7$ simulated AGN was initially calculated following the relation $\log\tau = K + \alpha(\log L - 44 ) + R(\sigma_{\rm int})$, where $R(\sigma_{\mathrm{int}})$ is a random number drawn from a normal distribution with a standard deviation of $\sigma_{\rm int}$. For a given luminosity, this process produced $\tau$ above or below the \cite{Bentz13} line. We designate this sample S1, shown in Figure \ref{simulation} as purple data points. Figure \ref{simulation} presents one iteration of the complete simulation.

\subsection{Observational Limits}

The SDSS-RM observational selection effects were applied to the simulations by adding observational uncertainties as well as lag and magnitude limits to the S1 sample. First, observational uncertainties were assigned to each of the simulated AGN by randomly drawing luminosity and lag uncertainties ($\sigma_L$ and $\sigma_\tau$) from the actual 44 SDSS-RM \Lc\ and $\tau$ measurements \citep{Shen15a,Grier17}. We then replicated the sample limits of SDSS-RM by imposing the same lag and magnitude constraints as the observations. Simulated AGN were restricted to observed-frame lags $4<\tau_{\rm obs}<75$~days, and $i$-band magnitude \less\ 21.7.

The average cadence for SDSS-RM observations was 4 days, which places a lower limit on the possible observed-frame time lags. Conversely the upper limit of 75 days comes from the longest measured time lag from SDSS-RM, related to the monitoring duration of 180 days and the need for overlap between the continuum and emission-line light curves.

While the observed-frame lag limit can be implemented by a simple redshift conversion, several additional steps were required to fully emulate the magnitude limits of the observed SDSS-RM sample. The SDSS-RM parent sample of quasars is restricted to total (AGN+host) magnitudes of $i<21.7$, but the S1 sample has AGN-only luminosities at rest-frame 5100 \AA. We add a host-galaxy contribution to the simulated AGN luminosities following Equation (\ref{Lest}) (measured for similar SDSS AGN spectra by \citealp{Shen11}). We assume a 0.35~dex scatter in this relation, since 0.35~dex is the standard deviation of the actual host-galaxy luminosities of the SDSS-RM quasars. \red{We convert this total \Lc\ to $i$-band magnitude before implementing a magnitude cutoff.} However, there is an additional magnitude dependence of the lag detection that must be considered, as lags are easier to recover for brighter AGN: the fraction of AGN from SDSS-RM with detected lags by \cite{Grier17} is roughly 1/3 as high for $i>20$ AGN as for $i<20$ AGN. We account for this by removing all AGN with $i>21.7$ and keeping all AGN with $i<20$, and only keeping 1/3 of AGN with $20<i<21.7$.

We designate this ``observation-limited" sample S2, shown as blue points in Figure \ref{simulation}. The boundaries in rest-frame lag and luminosity are smooth rather than sharp due to the range of redshifts applied to the simulated sample, and are slightly tilted because both the observed-frame lag and magnitude limits depend on redshift to convert to the rest-frame lag and luminosity. 

Finally, to account for the limit in the number of actual lag detections in SDSS-RM (44 measured lags), we randomly selected 44 points from S2; we designate this ``number-limited" sample S3. The S3 sample for one of the simulations is shown as the red points in Figure \ref{simulation}. 

\subsection{Fitting the Simulated $R-L$ relation}

\begin{figure}[!t]

\includegraphics[width=\columnwidth]{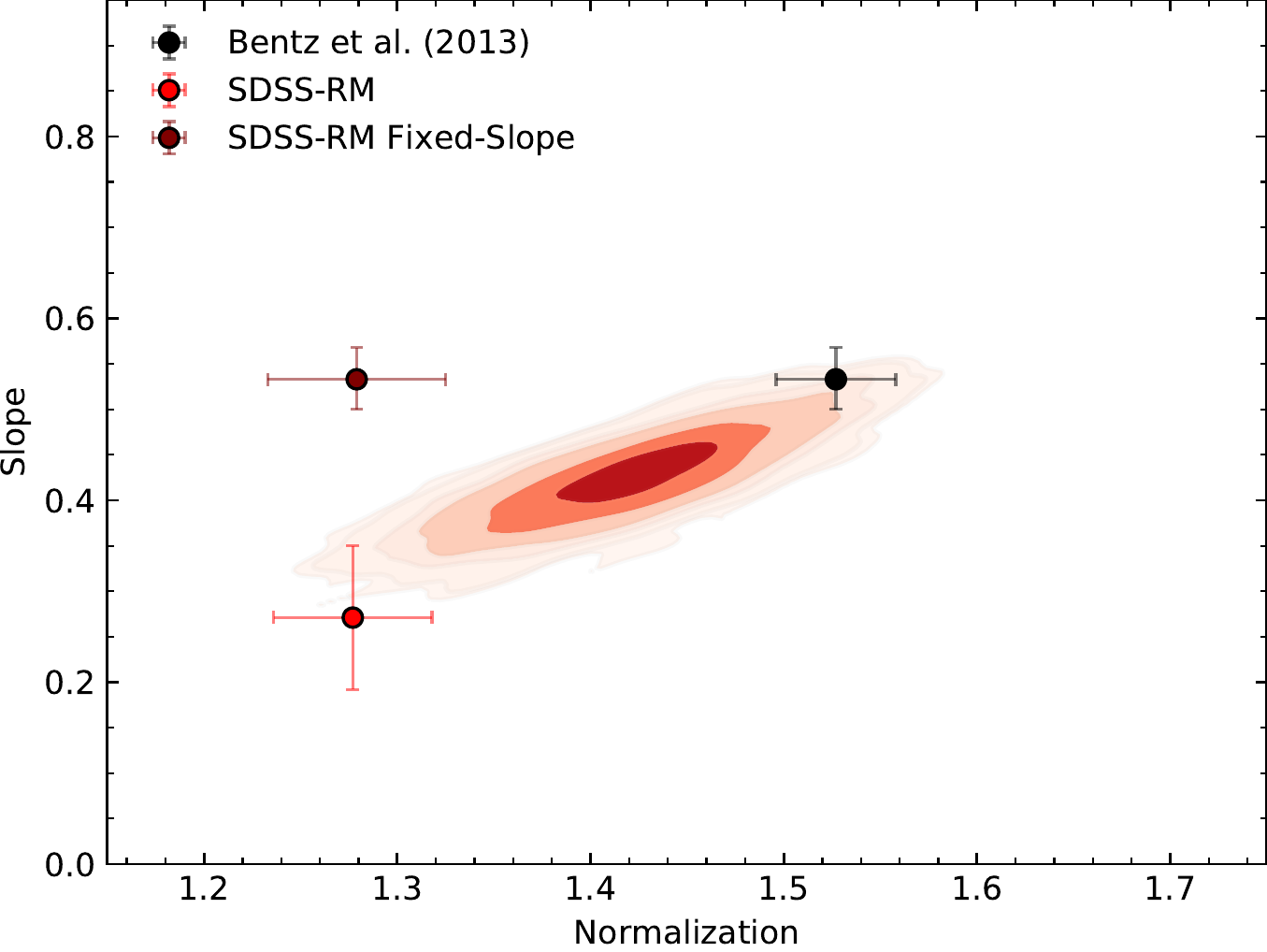}
\includegraphics[width=\columnwidth]{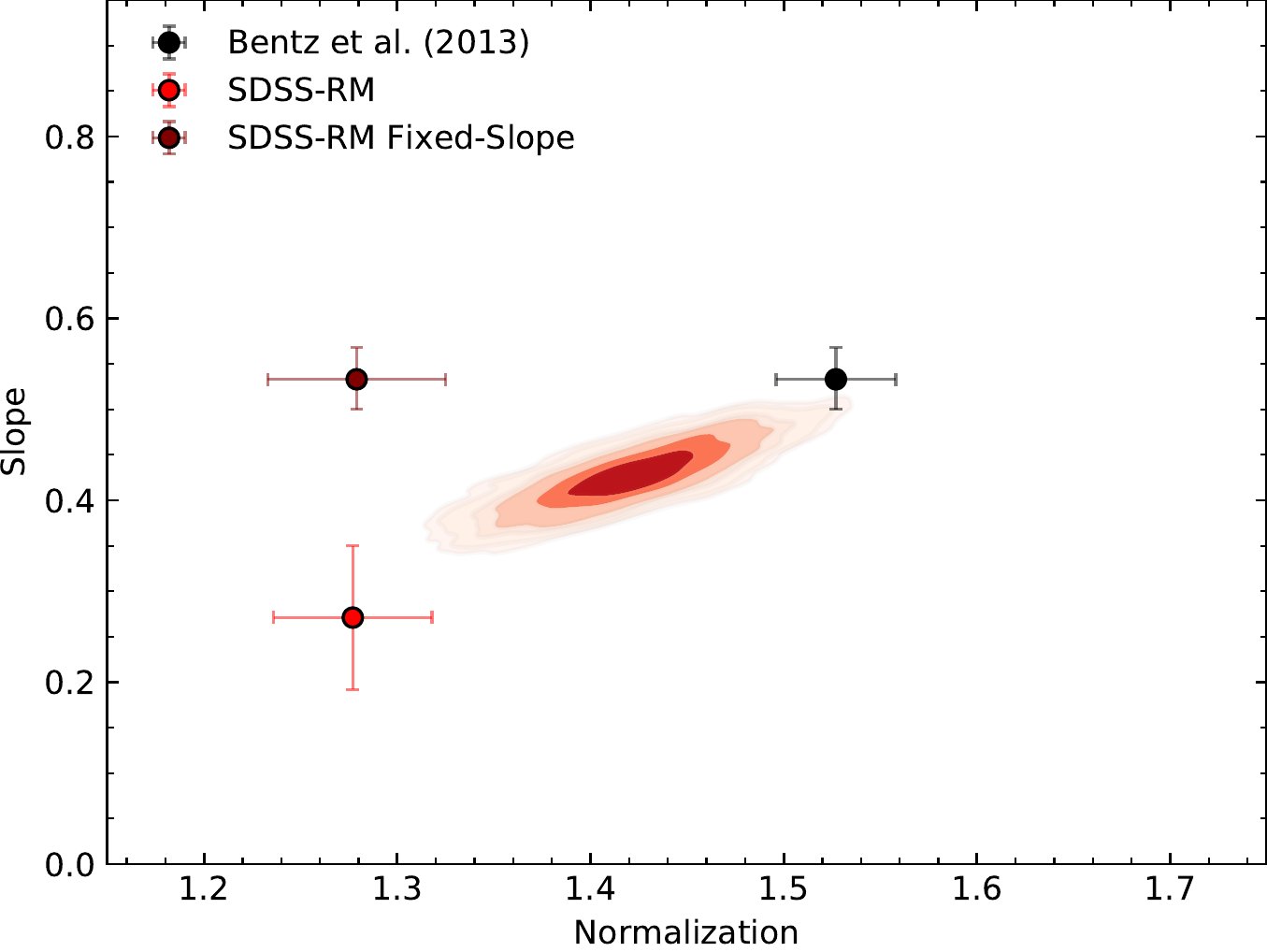}
\caption{\textit{Top:} The distribution of slopes and normalizations from fitting 44 random points from our simulated sample, shown as red contours that include \red{38\% (0.5$\sigma$),} 68\% (1$\sigma$), 86\% (1.5$\sigma$), 95\% (2$\sigma$), 98\% (2.5$\sigma$) and 99\% (3$\sigma$) of the distribution. The red point represents the fitting results for SDSS-RM (Figure \ref{sdssfit}). The black point represents the result from \cite{Bentz13}. The dark red point represents the fitting result for SDSS-RM keeping the slope fixed to be the same as \cite{Bentz13}. The SDSS-RM measurement falls outside the \red{3$\sigma$ contour and is only \less\ 1$\%$ likely to be produced by the simulation of observational bias. The \cite{Bentz13} measurement falls just outside the 2$\sigma$ contour and is consistent with 5\% of the simulated $R-L$ parameters.}
{\textit{Bottom:} The distribution of slopes and normalization for \red{100} random points from the simulated sample, using the same enclosed probabilities for the contour levels. \red{The SDSS-RM point is outside the 3$\sigma$ contour and so is again only $<1$\% likely to be consistent with the simulation}. In both cases, observational bias is insufficient to explain the $R-L$ offsets of the SDSS-RM quasars.}}
\label{slopecont}
\end{figure}

We repeated the random selection of 44 points and best-fit line \red{2000} times to see how observing specific AGN affected the slope of the simulated relation. We used the python package \red{\texttt{PyMC3}} to determine the best-fit $R-L$ relation for each of the \red{2000} simulations, with one example of this fit shown by the dashed red line in Figure \ref{simulation}. The distribution of best-fit line parameters from the \red{2000} simulations is presented in Figure \ref{slopecont}. The simulated best-fit $R-L$ relations have a median slope \red{$0.43^{+0.04}_{-0.04}$}, and a median normalization of \red{$1.42^{+0.04}_{-0.05}$}; here the plus and minus values represent the 16\% and 84\% percentiles of the distribution of slopes and normalizations, not the uncertainty in the fit. The slope and normalization are consistent \red{(2.6$\sigma$ and 2.7$\sigma$, respectively)} with the \cite{Bentz13} $R-L$ relation (represented by the black point in Figure \ref{slopecont}). Only \red{\less\ $1\%$} of the simulations have best-fit slopes and normalizations that are as extreme as the best-fit $R-L$ relation for the observed SDSS-RM data. This result suggests that observational biases are unlikely to be the main cause of the different $R-L$ relation represented by SDSS-RM AGN compared to previous RM samples.

\red{To examine if the number of detected lags by SDSS-RM affects the $R-L$ relation, we can increase the number of selected points to reflect the future number of detected \Hb lags. The Black Hole Mapper in the upcoming SDSS-V will allow RM of over 1000 quasars \citep{Kollmeier17}. We estimate that this will increase the number of \Hb\ lags to $\sim$ 100. Here we assume the SDSS-RM observational effects applied to the simulations are also a reasonable approximation for the SDSS-V observations. The distribution of best-fit lines for the 100 random points has a median slope \red{$0.43^{+0.03}_{-0.03}$} and normalization of \red{$1.42^{+0.03}_{-0.03}$}. Here the best-fit slope and normalization that are inconsistent \red{(by 3.5$\sigma$)} with the \cite{Bentz13} best-fit line, suggesting that a larger sample will better constrain the effects of observational bias.} The narrower distribution of best-fit lines is even less likely than the smaller simulated sample to match the observed SDSS-RM $R-L$ relation, with less than 1\% of the simulated best-fit $R-L$ relations as extreme as the best fit to the SDSS-RM observations.

%%%%%%%%%%%%%%%%%%

%%%%%%%%%%%%%%%%%%%%

Since slope and normalization are degenerate parameters in the best-fit $R-L$ relation, and considering the limited range in SDSS-RM luminosities, we additionally repeated the fitting procedure with slope fixed to the \cite{Bentz13} value of $\alpha = 0.533$ and only allowed the normalization $K$ to vary. This effectively tests if the simulations of observational bias can reproduce the $R-L$ offset of the SDSS-RM AGN. The mean normalization for the distribution is \red{$K = 1.53^{+0.03}_{-0.03}$, again consistent with $K = 1.527$ from \cite{Bentz13} and \more\ 5$\sigma$ inconsistent with the observed $R-L$ offset of the SDSS-RM data.}

In general the simulations of observational bias produce a $R-L$ relation that is statistically consistent with the \cite{Bentz13} best-fit relation, with only marginally flatter slopes and lower normalizations. \red{Less than 1\%} of the simulations produce best-fit $R-L$ relations that are as extreme as the observed SDSS-RM and SEAMBH $R-L$ data. \cite{Li19} arrived at a similar conclusion using independent light-curve simulations, additionally noting that JAVELIN lags measured from SDSS-RM data are unlikely to include enough false positive detections to strongly influence the measured $R-L$ relation.

Our simulations suggest that observational bias is unlikely to be the main cause of the SDSS-RM and SEAMBH AGN lags falling below the \cite{Bentz13} $R-L$ relation. In the next section we investigate the possibility that $R-L$ offsets are instead driven by physical AGN properties. 

%%%%%%%%%%%%%%%%%%%%%%%%%%%%%%%%%%%%%%%%%%%%%%%%%%%%%%%%%%%%%%%%%%%%%%%%%%%%%%%%%%%%%%
% Section 4
%%%%%%%%%%%%%%%%%%%%%%%%%%%%%%%%%%%%%%%%%%%%%%%%%%%%%%%%%%%%%%%%%%%%%%%%%%%%%%%%%%%%%%

\section{Properties of Quasars Offset from the $R-L$ relation}

The $R-L$ differences between SDSS-RM and \cite{Bentz13} may exist because the SDSS-RM sample spans a broader range of quasar properties \citep{Shen15a,Shen19}. The SEAMBH sample also occupies a very different parameter space compared to the \cite{Bentz13} sample, as SEAMBH AGN were specifically selected to have higher Eddington ratios. 
%\citet{Du16,Du18} used the SEAMBH sample to argue that, at fixed \Lc, $\tau$ inversely correlates with Eddington ratio. For the SDSS-RM AGN with lower accretion rates ($\dot{\mathcal{M}}$ \less\ 3), \cite{Du18} instead attributes the offset to retrograde accretion (i.e., SMBHs spinning counter to their accretion disks). It seems unlikely that SDSS-RM quasars are biased to be nearly all retrograde spinning black holes, since the sample was selected only by a magnitude limit and otherwise spans a broad range of quasar properties \citep{Shen15b,Shen19}.

In this section, we investigate how the offset from the \cite{Bentz13} $R-L$ relation depends on various AGN properties. We define this offset as the ratio between the measured rest-frame \Hb\ lag $\tau_{\mathrm{obs}}$ and the expected time lag $\tau_{R-L}$ from Equation (\ref{BRL}) for the given AGN \Lc. We calculate the offset (\tauoff) for each of the AGN in \cite{Grier17}, \cite{Bentz13}, and \cite{Du16,Du18}. \red{In the subsequent analyses, we report the significance of each correlation in terms of the factor of sigma by which its slope is inconsistent from zero, using 3$\sigma$ as our threshold for a significant correlation.}

\subsection{$R-L$ Offset with Accretion Rate}

%%%%%%%%%%%%%%%%%%

\begin{figure}[!t]
\centering
\includegraphics[width=\columnwidth]{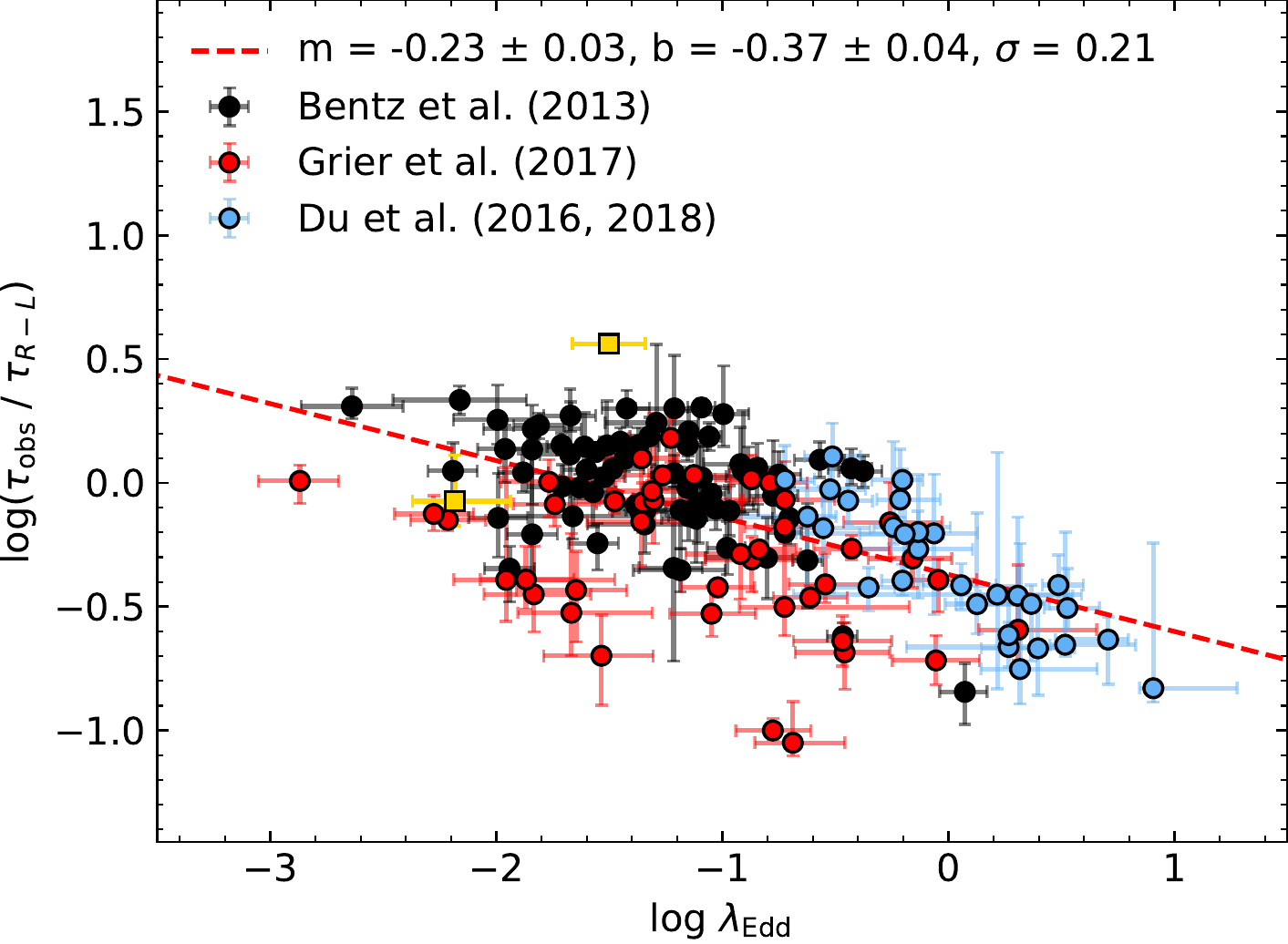}
\includegraphics[width=\columnwidth]{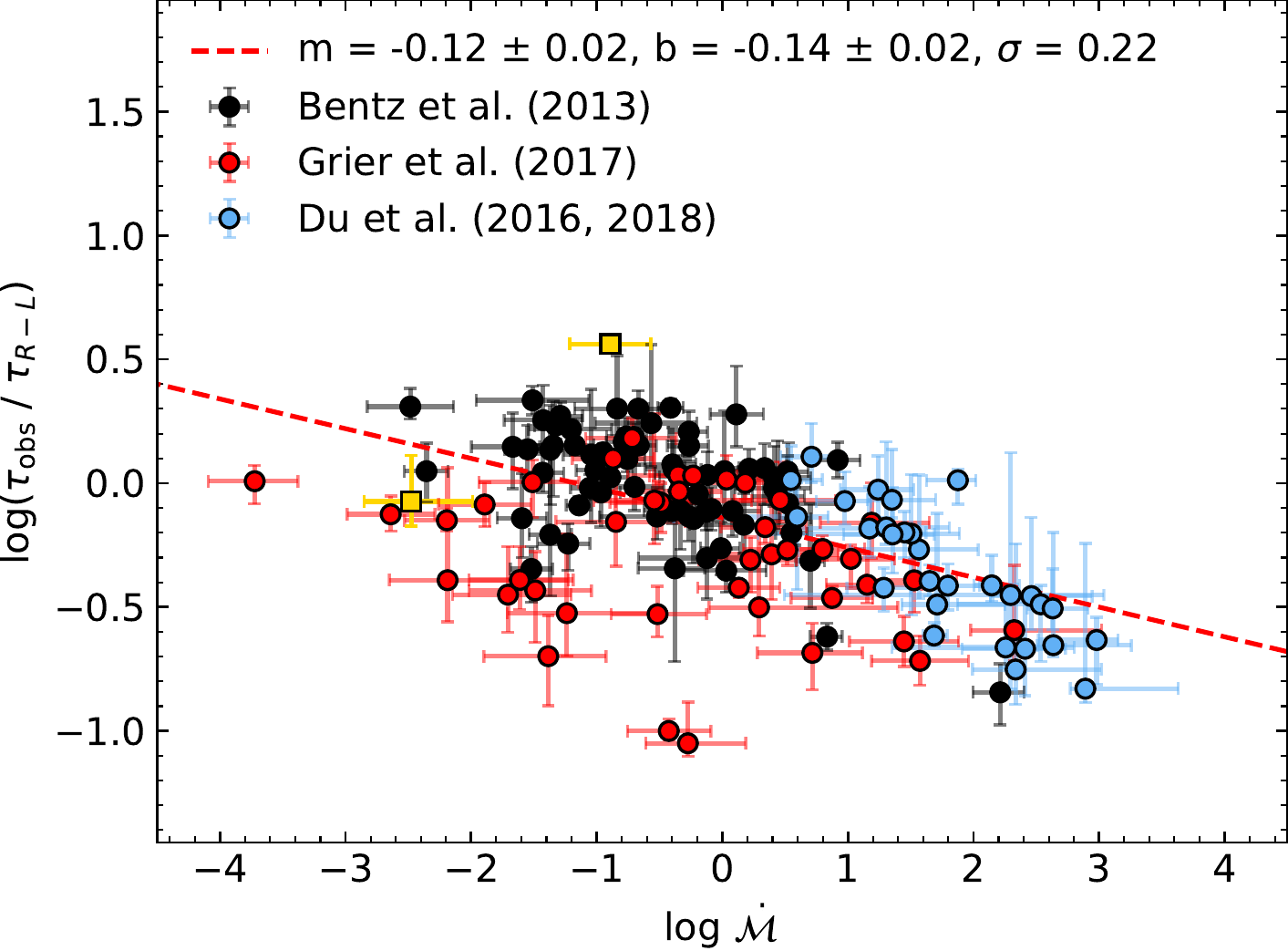}
\caption{The $R-L$ offset \tauoff\ of the three RM samples with Eddington ratio $\lambda_{\mathrm{Edd}}$ (top) and the pseudo accretion rate $\dot{\mathcal{M}}$ (see Equation \ref{mscript}). In both cases there is a significant anti-correlation between the two quantities, with the best-fit lines shown in red. The best-fit lines have slopes $m$ that are \more\ 5$\sigma$ different from zero, and a Spearman's $\rho \sim -0.50$ with a null-probability value of $p \sim 10^{-11}$. However, these trends are \red{difficult to interpret} since the two axes are self-correlated. We find much weaker correlations when comparing $R-L$ offsets to uncorrelated quantities associated with accretion rate, as seen in Figures \ref{LMdiff} and \ref{refdiff}.}
\label{eddingtondiff}
\end{figure}

%%%%%%%%%%%%%%%%%

\cite{Du16,Du18} propose that the $R-L$ offsets are driven by accretion rate, with more rapidly accreting AGN having shorter lags at fixed \Lc. They suggest that radiation pressure in rapidly accreting AGN causes the inner disk to be thicker (a ``slim'' disk), causing self-shadowing of the disk emission that reduces the ionizing radiation received by the BLR and thus decreases its radius \citep{Wang14c}. The self-shadowing does not affect the optical continuum emission used in the $R-L$ relation, so the broad-line lags are shorter than expected for a given observed \Lc. However, a correlation between offset and accretion rate is expected not just from quasar properties but simply because the axes are correlated: the y-axis (\tauoff) is a log-ratio of $\tau/\Lc^{0.5}$, while the x-axes ($\lambda_{\rm Edd}$, $\dot{\mathcal{M}}$) include log-ratios of \Lc/$\tau$ and $\Lc^{1.5}/\tau^2$, respectively. 

Despite these self-correlations, for direct comparisons to the previous SEAMBH results \citep[see][Figure 5]{Du18} we estimate accretion rates for all three samples using two dimensionless quantities: Eddington ratio \red{(calculated as described in Section 2) } and $\dot{\mathcal{M}}$ (Equation \ref{mscript}, as defined in \citealp{Du16}). The $R-L$ offsets of all three samples as a function of $\lambda_{\rm Edd}$ and $\dot{\mathcal{M}}$ are presented in Figure \ref{eddingtondiff}. Best-fit lines (with slope $m$ and y-intercept $b$ given in the figure legends) indicate significant (\more\ 5$\sigma$) anti-correlations between $R-L$ offset and both estimators of accretion rate, with Spearman's $\rho$ $\sim$ $-0.50$ and $p$ $\sim 10^{-11}$.

%%%%%%%%%%%%%%%%%%

\begin{figure}[!t]
\centering
\includegraphics[width=\columnwidth]{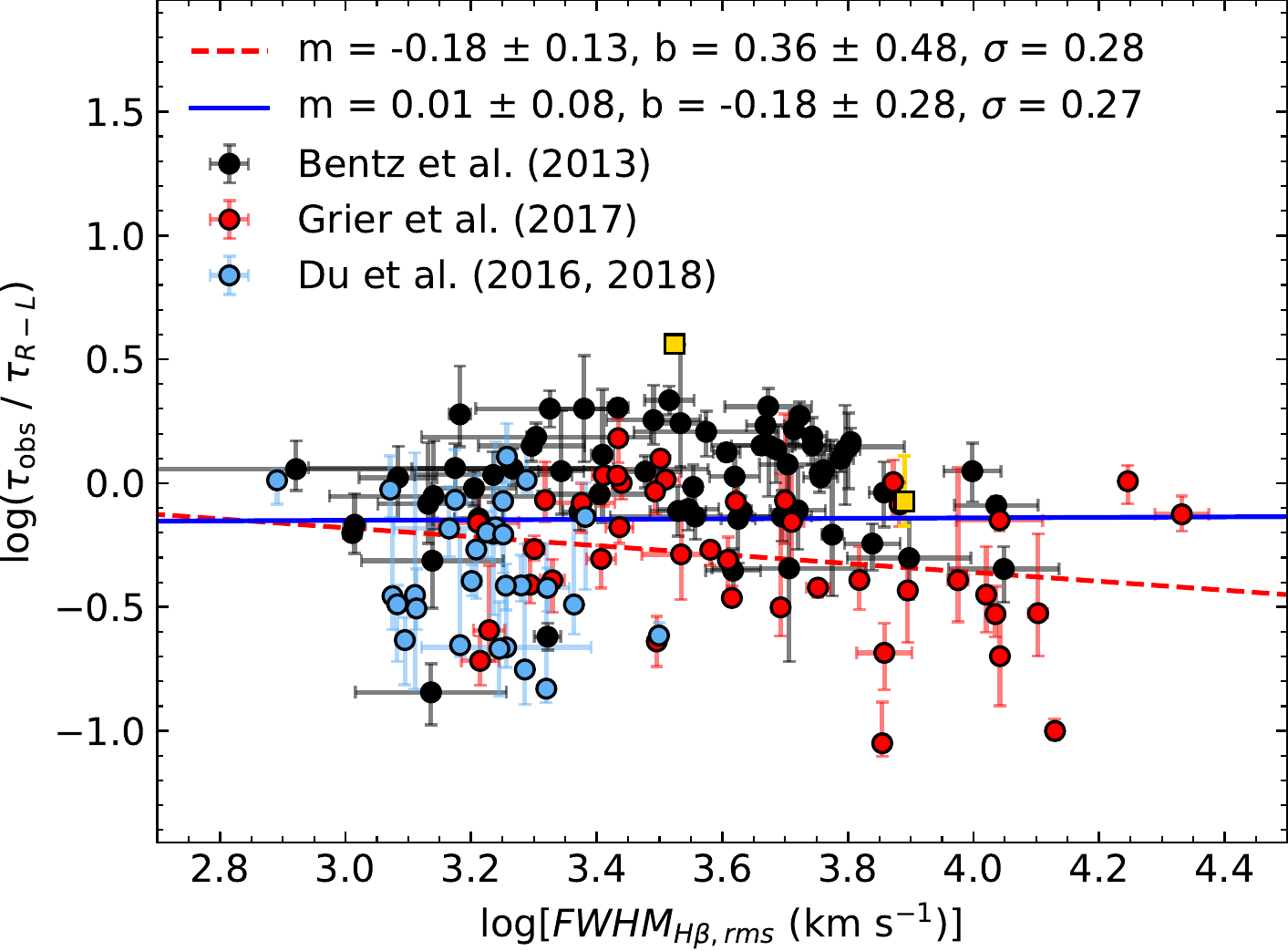}
\includegraphics[width=\columnwidth]{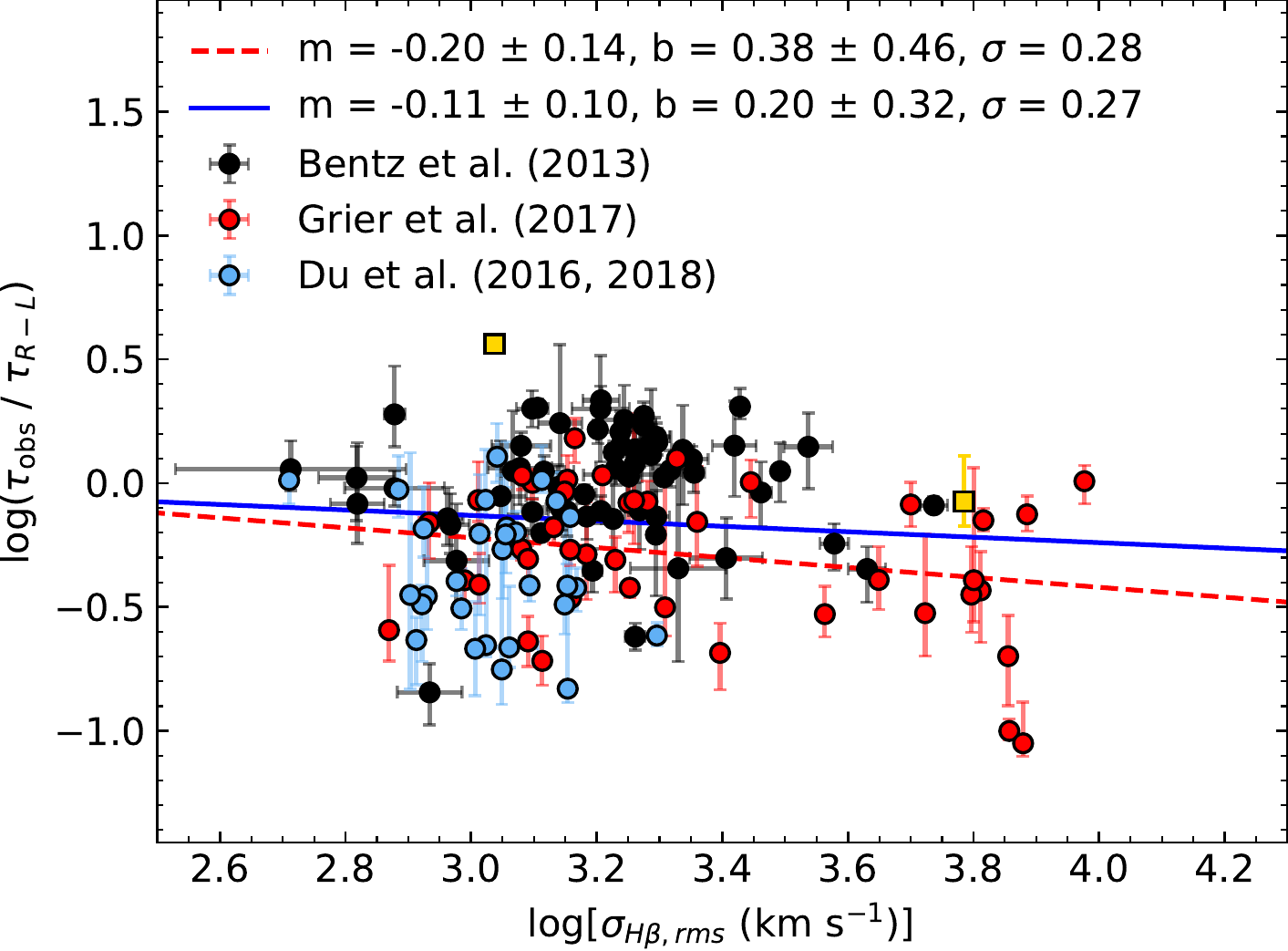}
\caption{The $R-L$ offset of AGN in all three samples with FWHM$_{\rm H\beta}$ (top panel) and $\sigma_{\rm H\beta}$ (bottom panel). \red{For the \cite{Bentz13} sample, the linewidths were taken from \cite{Bentz15}.} These observed quantities are related to Eddington ratio, and so are an attempt to connect $R-L$ offsets with accretion rate while avoiding direct self-correlation with $\tau$ on both axes. The red lines show the best-fit relations to the \cite{Grier17} SDSS-RM data, while the blue lines show the best-fit relations to all three samples. The $R-L$ offset is only marginally anti-correlated with the $\Hb$ line widths in each case.}

\label{LMdiff}
\end{figure}

%%%%%%%%%%%%%%%%%%%%%

The anti-correlations in both panels of Figure \ref{eddingtondiff} are qualitatively consistent with the simple self-correlations. To avoid these self-correlations, we instead study the dependence of $R-L$ offsets on accretion rate by using only the components of the Eddington ratio that are not computed directly from the the RM lag $\tau$. Since $\lambda_{\rm Edd} \equiv {L_{\rm bol}\over{L_{\rm Edd}}} \propto {{\Lc}\over{\Mbh}}$ and $\Mbh \propto \tau v_{\rm fwhm}^2$, we examine the $R-L$ offset against two measurements of line width $v_{\rm fwhm}$ and $v_{\rm \sigma}$ to determine if there are residual correlations beyond the self-correlations induced from $\Lc$ and $\tau$ appearing in both axes; this is presented in Figure \ref{LMdiff}. For all samples and for both line-width indicators, there are \red{only marginal (\less\ 2$\sigma$) anti-correlations between $R-L$ offset and $\Hb$ broad-line width.} 

%%%%%%%%%%%%%%%%%%

\begin{figure}[!t]
\centering
\includegraphics[width=\columnwidth]{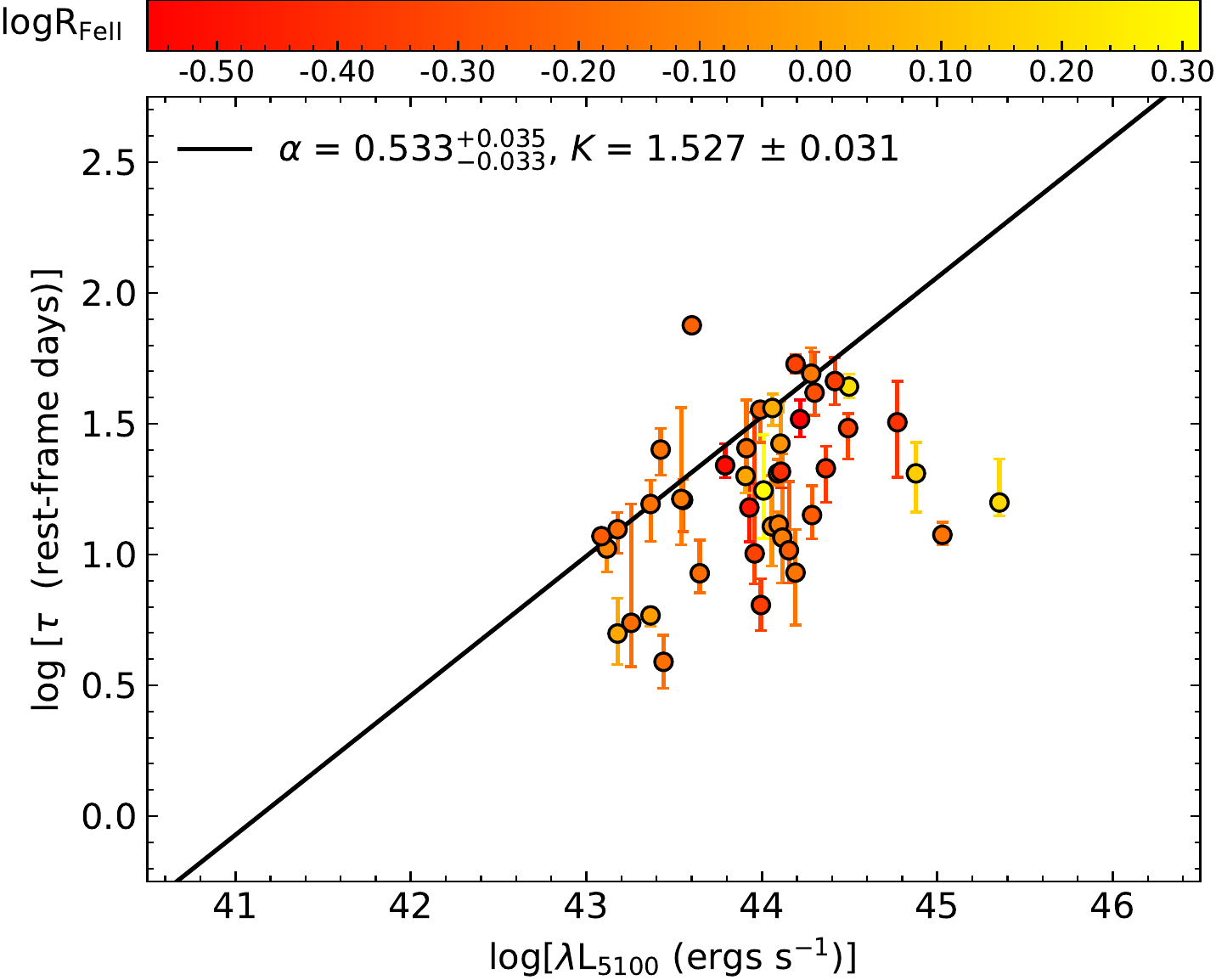}
\includegraphics[width=\columnwidth]{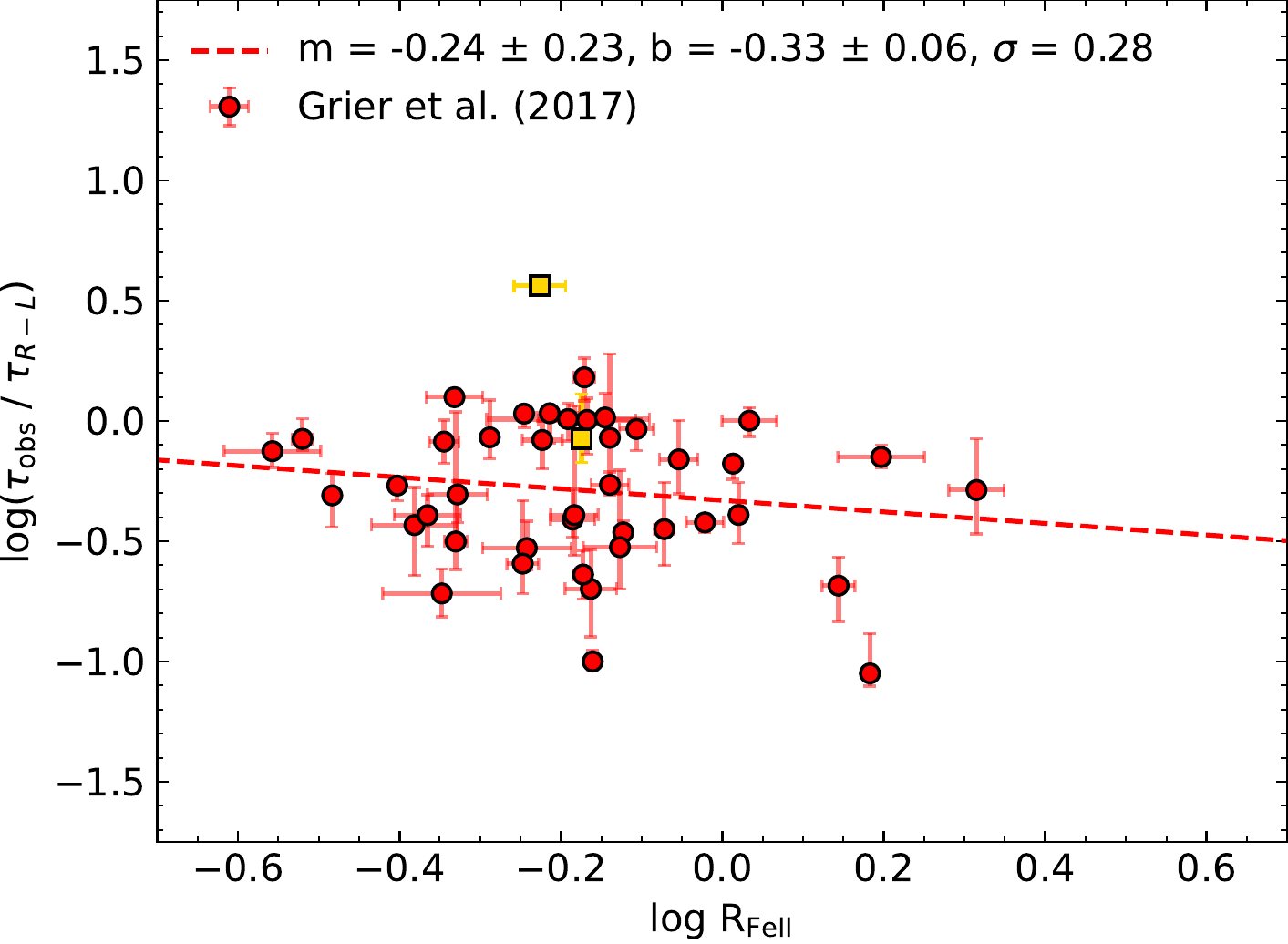}
\caption{The $R-L$ relation for SDSS-RM quasars color-coded by the \FeII\ effective strength $R_{\mathrm{FeII}}$ (top) and the $R-L$ offset \tauoff\ versus $R_{\mathrm{FeII}}$ (bottom). Since $R_{\mathrm{FeII}}$ correlates with Eddington ratio \citep{Shen14}, a significant anti-correlation between $R-L$ offset and $R_{\mathrm{FeII}}$ would suggest that, at fixed luminosity, more rapidly accreting AGN have shorter lags. We \red{do not observe a significant anti-correlation between $R-L$ offset and relative iron strength for SDSS-RM quasars, with a best-fit slope 1$\sigma$} consistent with zero, and Spearman's $\rho = -0.11$ and $p = 0.49$.}
\label{refdiff}
\end{figure}

%%%%%%%%%%%%%%%%%%%

We make a final attempt at studying the relation between $R-L$ offset and accretion rate by using the relative \FeII\ strength $R_{\mathrm{FeII}} \equiv \mathrm{EW_{FeII}\over{EW_{H\beta}}}$. The relative \FeII\ strength is one of the ``Eigenvector 1" quantities that separate quasars into different spectral categories \citep{Boroson92}, and in particular $R_{\mathrm{FeII}}$ correlates positively with Eddington ratio \citep{Shen14}. Thus we can use $R_{\mathrm{FeII}}$ as an independent estimate of accretion rate that avoids any self-correlation with \tauoff. Figure \ref{refdiff} presents the relation between $R-L$ offset and $R_{\mathrm{FeII}}$ for the SDSS-RM AGN of \cite{Grier17}. We find \red{no anti-correlation between offset and $\mathrm{R_{{FeII}}}$, with a slope that is 1$\sigma$ consistent with zero} and Spearman's $\rho = -0.11$ and $p = 0.49$. This is in contrast to the recent work of \cite{Du19}, who found a significant correlation between $R-L$ offset and $\mathrm{R_{{FeII}}}$ using the SEAMBH and \citet{Bentz13} AGN samples. We do find a consistent slope in the relation (\red{$m=-0.24 \pm 0.23$} compared to $m=-0.42 \pm 0.06$ in \citealp{Du19}), and our anti-correlation may be marginal rather than significant due to the limited sample size of SDSS-RM, the different lag uncertainties of JAVELIN, and/or the greater diversity of AGN properties in the SDSS-RM sample.

\subsection{$R-L$ Offset with UV Ionizing Luminosity}

The $R-L$ relation is parameterized with the optical luminosity at rest-frame 5100 \AA, but the response and size of the BLR is governed by the incident ionizing photons \citep[e.g,][]{Davidson72}. In particular, the $\Hb$ recombination line is driven by the incident luminosity of $E>13.6$~eV photons. The basic photoionization expectation of $R$ $\propto L^{0.5}$ is valid for the optical luminosity only if changes in optical luminosity also correspond to identical changes in the ionizing luminosity\red{; however, the shape of the SED and therefore the ratio of UV and optical luminosities depends on factors like mass, accretion rate and spin \citep[e.g.,][]{Richards06}. Modeling of the BLR by \cite{Czerny19}, with the assumption that the BLR radius is determined by the ionizing luminosity or number of incident photos, shows a diversity in the R-L relation due to changing UV/optical luminosity ratios, reproducing the range of observed lags from RM surveys. Additionally, emission-line lags relative to the optical continuum may be underestimated if there is an additional time lag between the UV-continuum and optical-continuum variability, as observed in NGC 5548  \citep{Pei17}. However, lags between the UV and optical continuum are short, so this would only significantly affect emission-line lags in the order of a few days.}

\begin{figure*}[!t]
\centering
\includegraphics[width=\columnwidth]{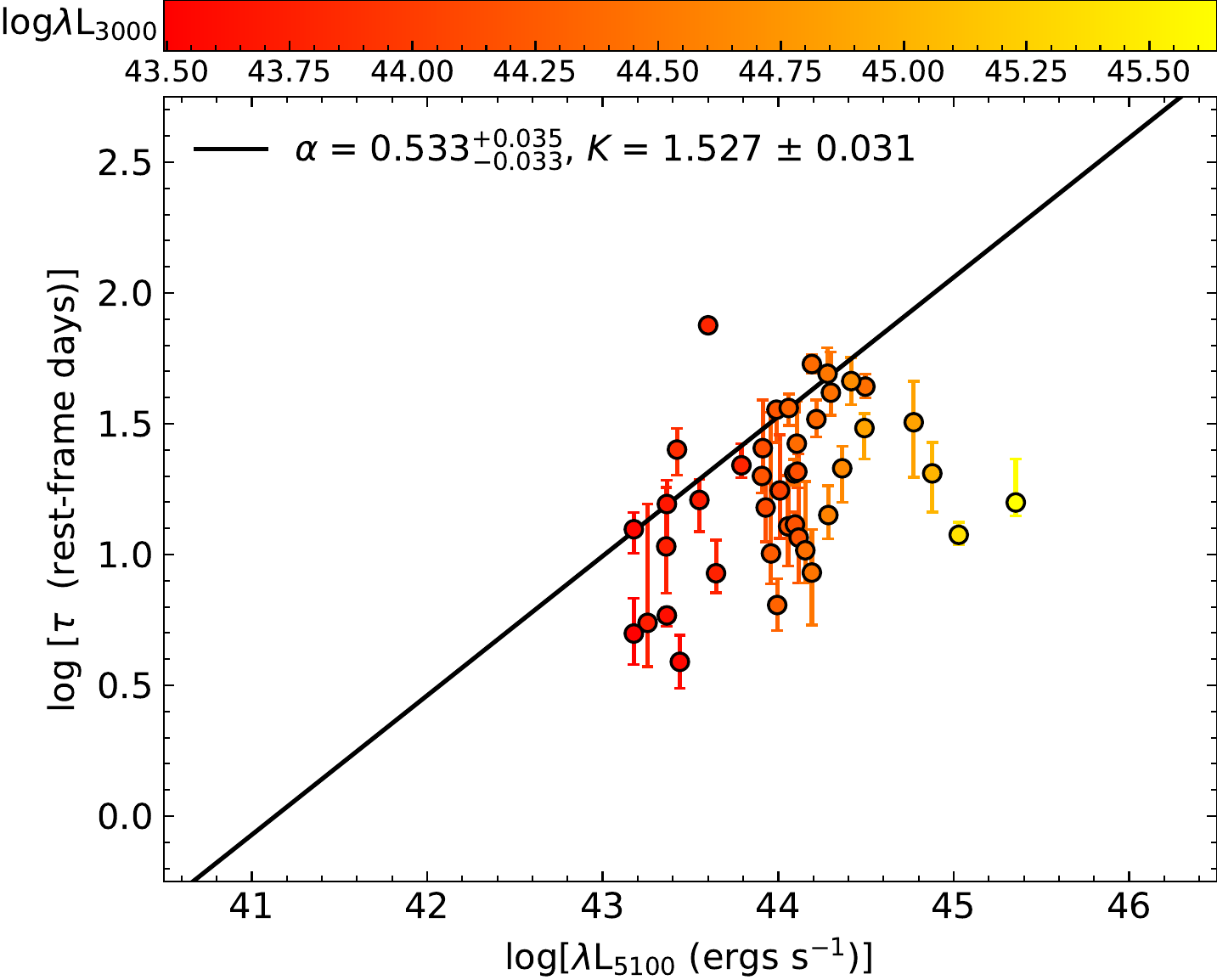}
\includegraphics[width=\columnwidth]{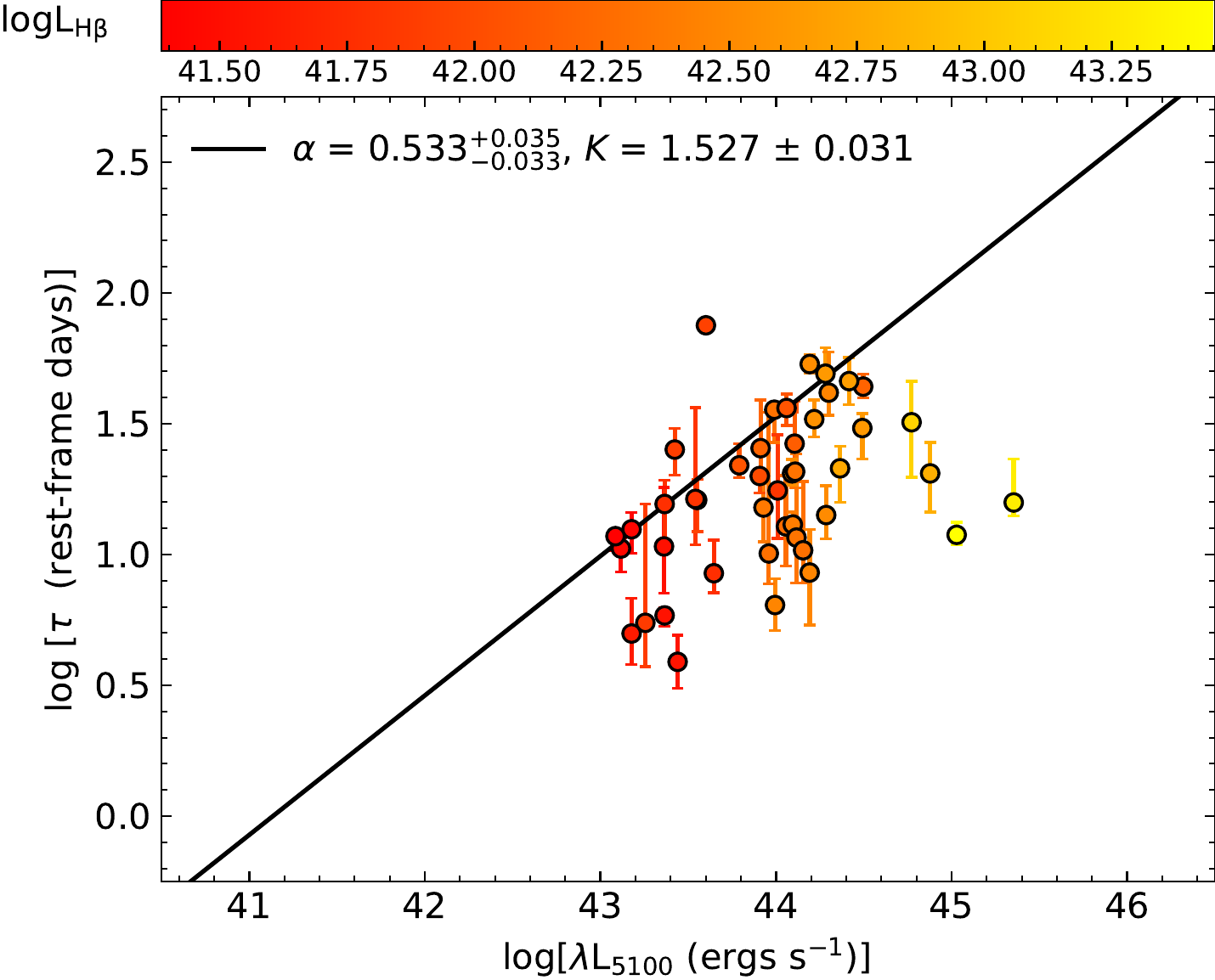}
\includegraphics[width=\columnwidth]{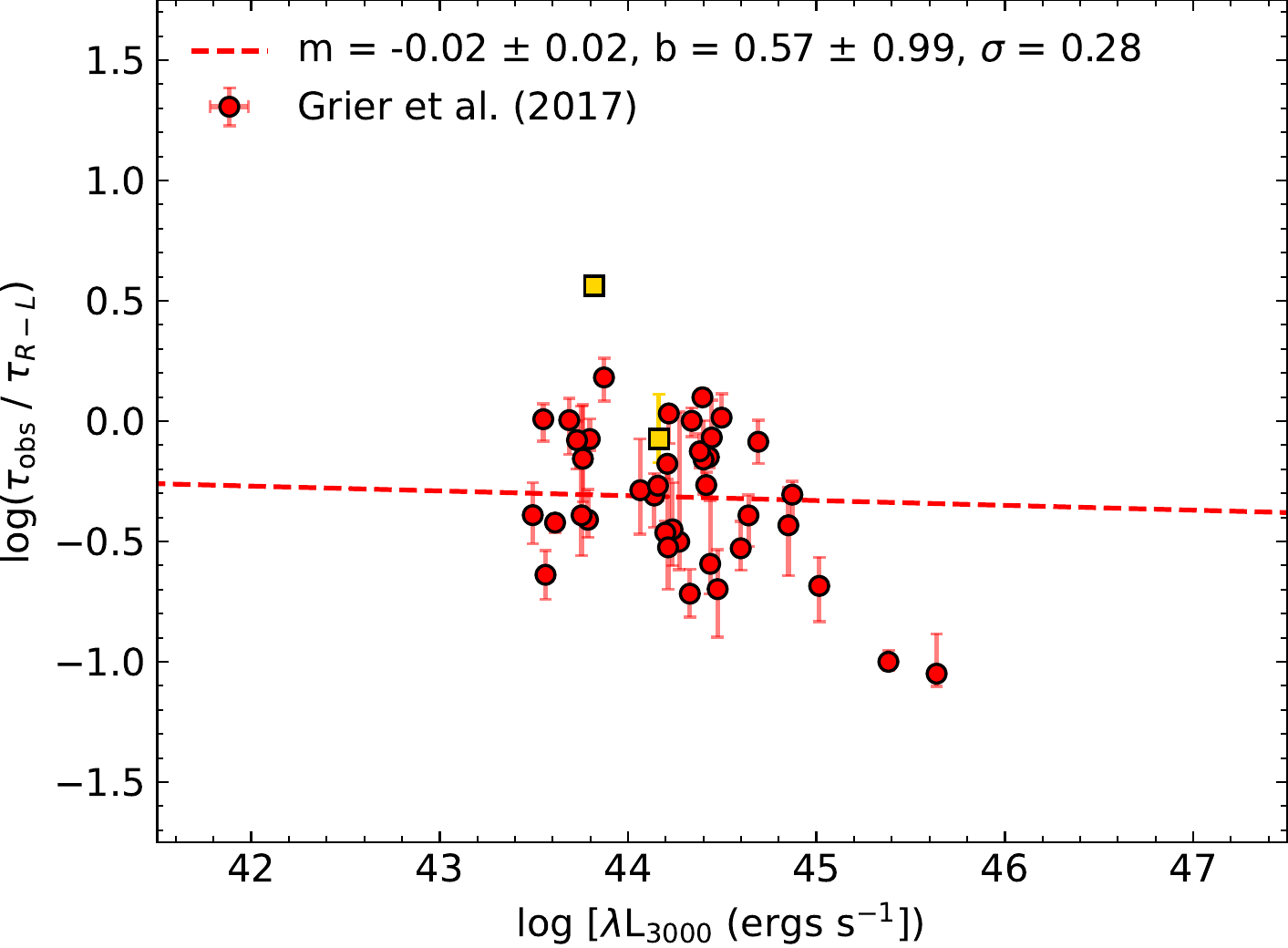}
\includegraphics[width=\columnwidth]{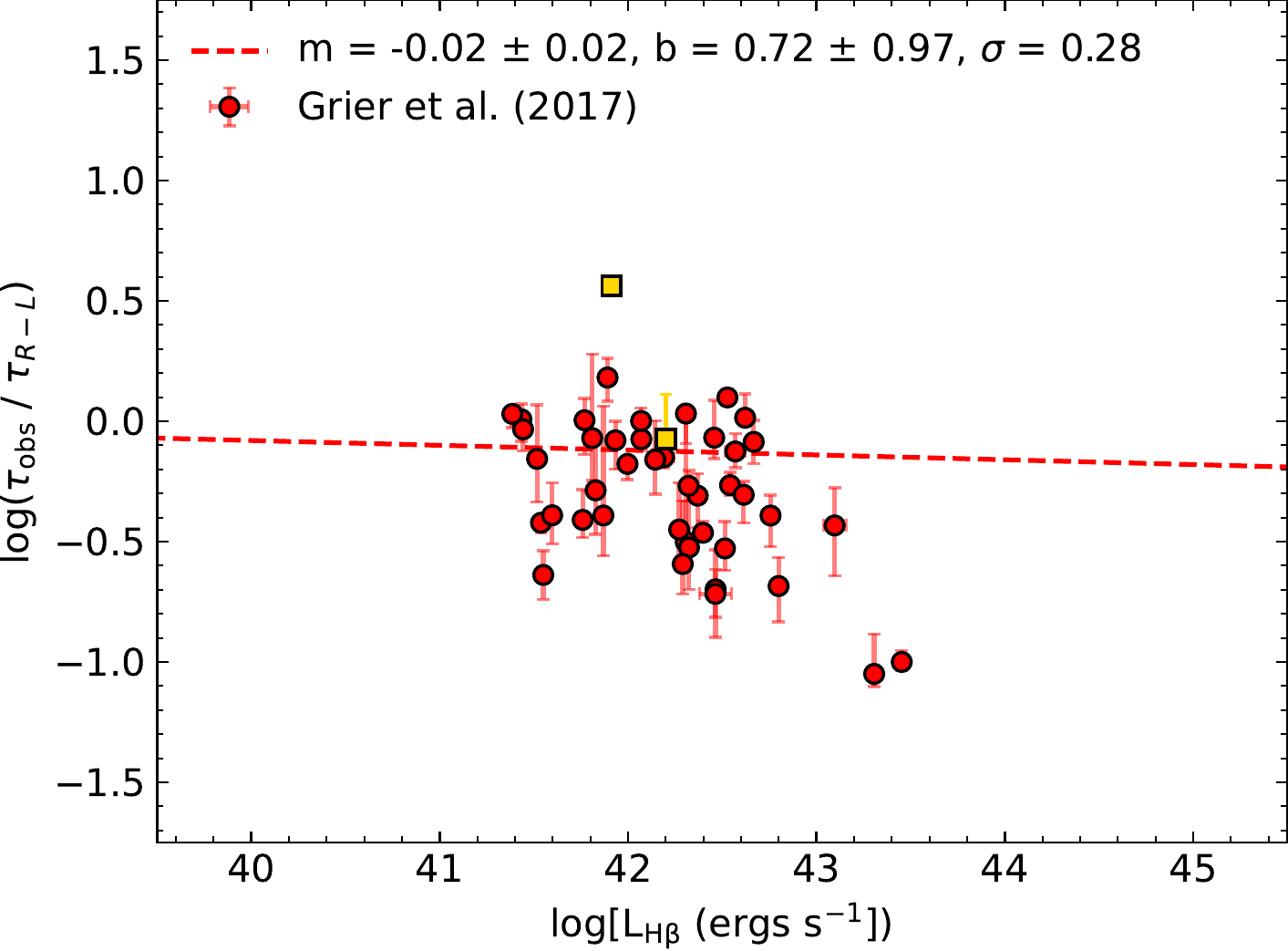}
\caption{\textit{Left:} The $R-L$ relation of SDSS-RM AGN color-coded by \Luv\ and the $R-L$ offset \tauoff\ versus \Luv, a luminosity measurement closer to the ionizing UV luminosity than the $\Lc$ used in the $R-L$ relation. The sample spans a fairly narrow range of \Luv/\Lc\ (top left) and the $R-L$ offset is \red{not anti-correlated with \Luv\ (slope 1$\sigma$ from zero and Spearman's $\rho = -0.28$ and $p = 0.09$}). \textit{Right:} The $R-L$ relation of SDSS-RM AGN with the $\Hb$ broad-line luminosity and the $R-L$ offset versus the $\Hb$ broad-line luminosity, a proxy for the ionizing luminosity that drives $\Hb$ recombination. Once more the sample spans a fairly narrow range of $L_{\rm H\beta}/\Lc$, \red{and} the $R-L$ offset and $L_{\rm H\beta}$ are \red{not significantly correlated with a slope $m$ that is 1$\sigma$ consistent with zero}, and Spearman's $\rho = -0.36$ and $p=0.02$, with excess scatter of $\sim$0.25~dex about the best-fit line.}
\label{iondiff}
\end{figure*}

None of our samples have published measurements of the $E>13.6$~eV ionizing luminosity; however, the SDSS-RM sample has luminosity measurements at rest-frame 3000~\AA\ and \Hb, better probing the (near-)UV compared to the optical \Lc. Both of these quantities are shown with the $R-L$ offset of SDSS-RM AGN in Figure \ref{iondiff}. We fit lines to each, finding that there is \red{no anti-correlation between the $R-L$ offset and \Luv, with Spearman's $\rho = -0.28$, $p = 0.09$.} The best-fit line finds \red{no significant (1$\sigma$) anti-correlation} between the $R-L$ offset and the $\Hb$ luminosity with Spearman's $\rho = -0.36$ and $p = 0.02$, \red{additionally} the $R-L$ relation color-coded by $L_{\rm H\beta}$ (Figure \ref{iondiff} top right) indicates little variation of $L_{\rm H\beta}/\Lc$ across the SDSS-RM sample.

\begin{figure}[!t]
\centering
\includegraphics[width=\columnwidth]{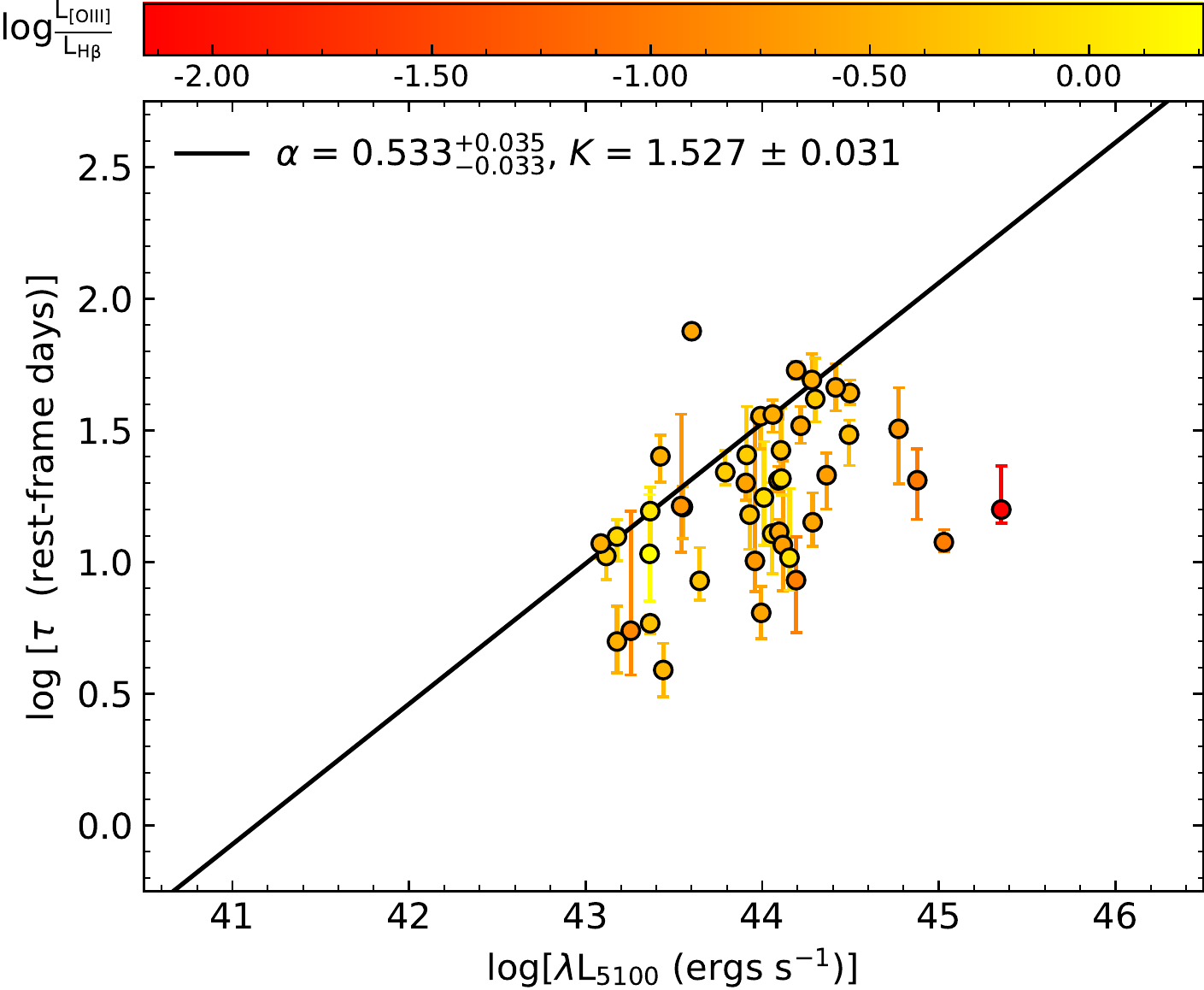}
\includegraphics[width=\columnwidth]{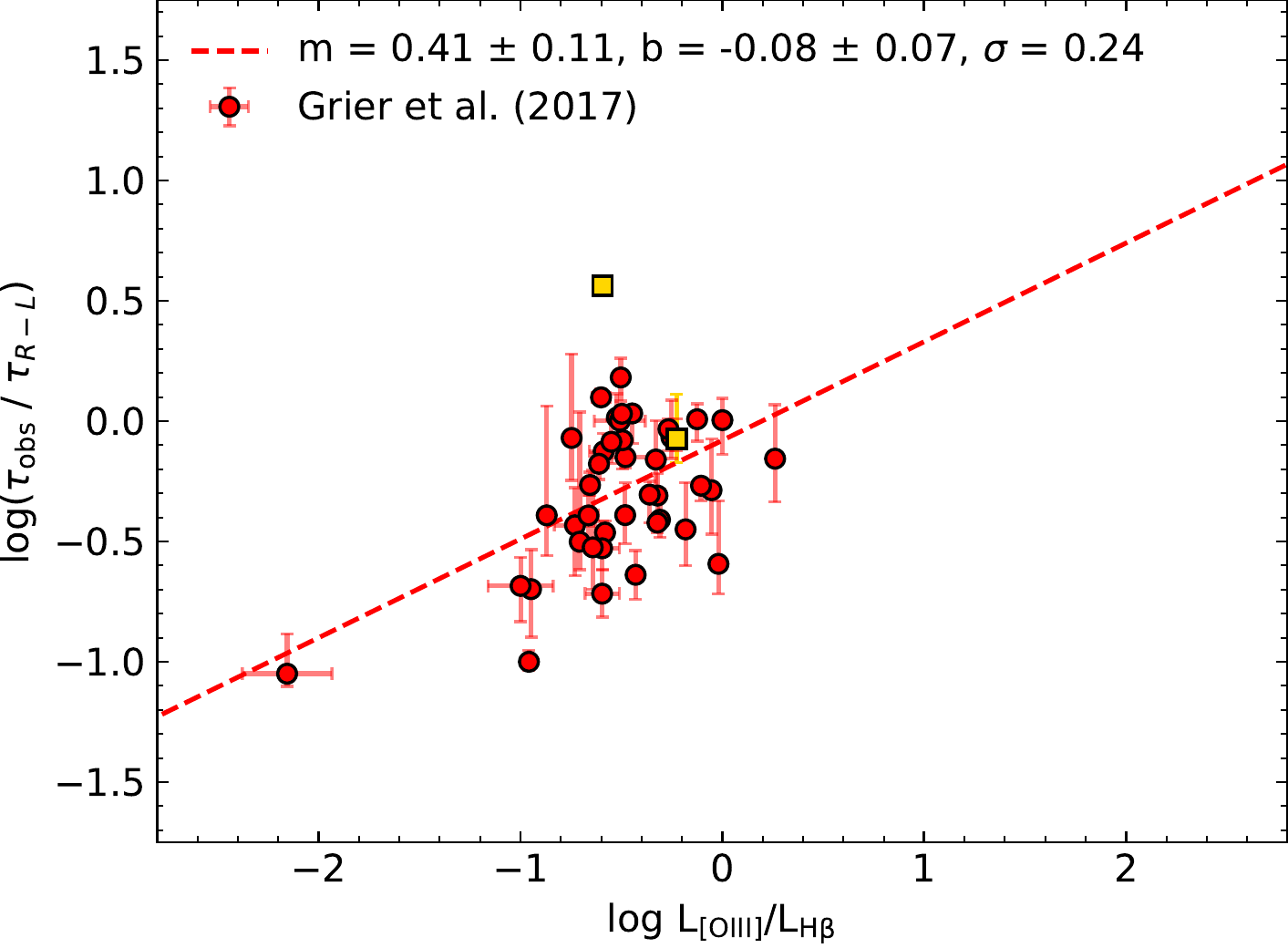}
\caption{The $R-L$ relation of SDSS-RM color-coded by $L_{\rm [OIII]}/{L_{\rm H\beta}}$ (top) and the $R-L$ offset \tauoff\ with $L_{\rm [OIII]}/{L_{\rm H\beta}}$ (bottom), an indicator of the far-UV ionizing flux present. AGN with larger (negative) $R-L$ offsets typically have lower $L_{\rm [OIII]}/{L_{\rm H\beta}}$, and there is a significant correlation between \tauoff\ and $L_{\rm [OIII]}/{L_{\rm H\beta}}$ with a slope \red{3.7$\sigma$} different from zero and a Spearman's $\rho = 0.36$ and $p = 0.02.$} 
\label{OIIIdiff}
\end{figure}

The ratio of luminosities of the $\OIII\lambda$5007 and \Hb\ emission lines is also frequently used as a proxy for the number of ionizing photons (e.g., \citealp{BPT81,VO87}). Both are recombination lines, and $\OIII$ has an ionization energy of 55~eV compared to the H ionization energy of 13.6~eV. \red{We find a significant (3.7$\sigma$) correlation between offset and L$_{\rm [OIII]}$/L$_{\rm H\beta}$, shown in Figure \ref{OIIIdiff}, with Spearman's $\rho = 0.36$, $p = 0.02$, and an excess scatter of $\sim$ 0.24.}

%%%%%%%%%%%%%%%%%%%

%%%%%%%%%%%%%%%%%%

We conclude that the shape of the UV/optical SED is likely to play a role in the $R-L$ offset of AGN, as evident from the correlation with $L_{\rm [OIII]}/L_{\rm H\beta}$. The \red{lack of significant} correlations with \Luv\ and $L_{\rm H\beta}$ may be because these luminosities do not accurately represent the luminosity of far-UV ($\lambda<912$ \AA) ionizing photons. The $L_{\rm [OIII]}/L_{\rm H\beta}$ ratio is likely tied to the broader shape of the AGN SED, which in turn is related to the accretion rate and/or black hole spin \red{\citep[e.g.,][]{Du18,Czerny19}}. It is a bit surprising that we find a significant correlation of $R-L$ offset with $L_{\rm [OIII]}/L_{\rm H\beta}$ but only a marginal anti-correlation with $R_{\mathrm{FeII}}$, given the observed anti-correlation between $\OIII$ equivalent width and $R_{\mathrm{FeII}}$ \citep[Figure 1 of][]{Shen14}. This may be due to the limited sample size of SDSS-RM AGN, and/or to the large uncertainties in its measured lags. Regardless of the root cause of the UV/optical SED changes, it would be valuable to add far-UV observations to the samples of SDSS-RM and SEAMBH AGN in order to directly compare their $R-L$ offsets with the luminosity of photons responsible for ionizing the BLR.

%%%%%%%%%%%%%%%%%%%%%%%%%%%%%%%%%%%%%%%%%%%%%%%%%%%%%%%%%%%%%%%%%%%%%%%%%%%%%%%%%%%%%%
% Section 5
%%%%%%%%%%%%%%%%%%%%%%%%%%%%%%%%%%%%%%%%%%%%%%%%%%%%%%%%%%%%%%%%%%%%%%%%%%%%%%%%%%%%%%

\section{Conclusions}

While previous RM studies revealed a tight ``$R-L$'' relation between the broad-line radius $R=c\tau$ and the optical luminosity \Lc, more recent studies (SDSS-RM and SEAMBH) frequently find shorter lags than expected for a given optical luminosity. We use Monte Carlo simulations that mimic the SDSS-RM survey design to show that the $R-L$ offsets are not solely due to observational bias. Instead, we find that AGN $R-L$ properties correlate most closely with AGN spectral properties: at fixed \Lc, AGN have lower $\tau$ with lower L$_{\rm [OIII]}$/L$_{\rm H\beta}$. The correlation of $R-L$ offset with L$_{\rm [OIII]}$/L$_{\rm H\beta}$ is likely tied to changes in the UV/optical spectral shape. A more complete understanding of AGN $R-L$ properties will likely come from observations of the UV SED of RM AGN that directly measure the luminosity and shape of the ionizing continuum responsible for the AGN broad-line region.

%%%%%%%%%%%%%%%%%%%%%%%%%%%%%%%%%%%%%%%%%%%%%%%%%%%%%%%%%%%%%%%%%%%%%%%%%%%%%%%%%%%%%%
% Acknowledgements
%%%%%%%%%%%%%%%%%%%%%%%%%%%%%%%%%%%%%%%%%%%%%%%%%%%%%%%%%%%%%%%%%%%%%%%%%%%%%%%%%%%%%%

\acknowledgments

LCH acknowledges support from the National Science Foundation of China (11721303, 11991052) and the National Key R\&D Program of China (2016YFA0400702). GFA, JRT, and YH acknowledge support from NASA grants HST-GO-15260.001-A and HST-GO-15650.002-A. YS acknowledges support from an Alfred P. Sloan Research Fellowship and NSF grant AST-1715579. KH acknowledges support from STFC grant ST/R000824/1. Funding for SDSS-III was provided by the Alfred P. Sloan Foundation, the Participating Institutions, the National Science Foundation, and the U.S. Department of Energy Office of Science. The SDSS-III web site is http://www.sdss3.org/.

%%%%%%%%%%%%%%%%%%

%%%%%%%%%%%%%%%%%%%%

SDSS-III was managed by the Astrophysical Research Consortium for the Participating Institutions of the SDSS-III Collaboration including the University of Arizona, the Brazilian Participation Group, Brookhaven National Laboratory, Carnegie Mellon University, University of Florida, the French Participation Group, the German Participation Group, Harvard University, the Instituto de Astrofisica de Canarias, the Michigan State/Notre Dame/JINA Participation Group, Johns Hopkins University, Lawrence Berkeley National Laboratory, Max Planck Institute for Astrophysics, Max Planck Institute for Extraterrestrial Physics, New Mexico State University, New York University, Ohio State University, Pennsylvania State University, University of Portsmouth, Princeton University, the Spanish Participation Group, University of Tokyo, University of Utah, Vanderbilt University, University of Virginia, University of Washington, and Yale University.

This work is based on observations obtained with MegaPrime/MegaCam, a joint project of CFHT and CEA/DAPNIA, at the Canada-France-Hawaii Telescope (CFHT) which is operated by the National Research Council (NRC) of Canada, the Institut National des Sciences de l’Univers of the Centre National de la Recherche Scientifique of France, and the University of Hawaii. The authors recognize the cultural importance of the summit of Maunakea to a broad cross section of the Native Hawaiian community. The astronomical community is most fortunate to have the opportunity to conduct observations from this mountain.

We thank the Bok and CFHT Canadian, Chinese, and French TACs for their support. This research uses data obtained through the Telescope Access Program (TAP), which is funded by the National Astronomical Observatories, Chinese Academy of Sciences, and the Special Fund for Astronomy from the Ministry of Finance in China.

\bibliography{RL}

\begin{thebibliography}{}
\expandafter\ifx\csname natexlab\endcsname\relax\def\natexlab#1{#1}\fi

\bibitem[{{Baldwin} {et~al.}(1981){Baldwin}, {Phillips}, \&
  {Terlevich}}]{BPT81}
{Baldwin}, J.~A., {Phillips}, M.~M., \& {Terlevich}, R. 1981, \pasp, 93, 5

\bibitem[{{Bentz} \& {Katz}(2015)}]{Bentz15}
{Bentz}, M.~C., \& {Katz}, S. 2015, \pasp, 127, 67

\bibitem[{{Bentz} {et~al.}(2009){Bentz}, {Walsh}, {Barth}, {Baliber},
  {Bennert}, {Canalizo}, {Filippenko}, {Ganeshalingam}, {Gates}, {Greene},
  {Hidas}, {Hiner}, {Lee}, {Li}, {Malkan}, {Minezaki}, {Sakata}, {Serduke},
  {Silverman}, {Steele}, {Stern}, {Street}, {Thornton}, {Treu}, {Wang}, {Woo},
  \& {Yoshii}}]{Bentz09}
{Bentz}, M.~C., {Walsh}, J.~L., {Barth}, A.~J., {et~al.} 2009, \apj, 705, 199

\bibitem[{Bentz {et~al.}(2013)Bentz, Denney, Grier, Barth, Peterson,
  Vestergaard, Bennert, Canalizo, Rosa, Filippenko, Gates, Greene, Li, Malkan,
  Pogge, Stern, Treu, \& Woo}]{Bentz13}
Bentz, M.~C., Denney, K.~D., Grier, C.~J., {et~al.} 2013, The Astrophysical
  Journal, 767, 149

\bibitem[{{Blandford} \& {McKee}(1982)}]{Mckee}
{Blandford}, R.~D., \& {McKee}, C.~F. 1982, \apj, 255, 419

\bibitem[{{Boroson} \& {Green}(1992)}]{Boroson92}
{Boroson}, T.~A., \& {Green}, R.~F. 1992, \apjs, 80, 109

\bibitem[{{Collin} {et~al.}(2006){Collin}, {Kawaguchi}, {Peterson}, \&
  {Vestergaard}}]{Collin06}
{Collin}, S., {Kawaguchi}, T., {Peterson}, B.~M., \& {Vestergaard}, M. 2006,
  \aap, 456, 75

\bibitem[{{Czerny} {et~al.}(2019){Czerny}, {Wang}, {Du}, {Hryniewicz}, {Karas},
  {Li}, {Pand a}, {Sniegowska}, {Wildy}, \& {Yuan}}]{Czerny19}
{Czerny}, B., {Wang}, J.-M., {Du}, P., {et~al.} 2019, \apj, 870, 84

\bibitem[{{Davidson}(1972)}]{Davidson72}
{Davidson}, K. 1972, \apj, 171, 213

\bibitem[{{Dexter} {et~al.}(2019){Dexter}, {Xin}, {Shen}, {Grier}, {Liu},
  {Gezari}, {McGreer}, {Brand t}, {Hall}, {Horne}, {Simm}, {Merloni}, {Green},
  {Vivek}, {Trump}, {Homayouni}, {Peterson}, {Schneider}, {Kinemuchi}, {Pan},
  \& {Bizyaev}}]{Dexter19}
{Dexter}, J., {Xin}, S., {Shen}, Y., {et~al.} 2019, arXiv e-prints,
  arXiv:1906.10138

\bibitem[{{Du} \& {Wang}(2019)}]{Du19}
{Du}, P., \& {Wang}, J.-M. 2019, arXiv e-prints, arXiv:1909.06735

\bibitem[{{Du} {et~al.}(2016){Du}, {Lu}, {Zhang}, {Huang}, {Wang}, {Hu}, {Qiu},
  {Li}, {Fan}, {Fang}, {Bai}, {Bian}, {Yuan}, {Ho}, {Wang}, \& {SEAMBH
  Collaboration}}]{Du16}
{Du}, P., {Lu}, K.-X., {Zhang}, Z.-X., {et~al.} 2016, \apj, 825, 126

\bibitem[{{Du} {et~al.}(2018){Du}, {Zhang}, {Wang}, {Huang}, {Zhang}, {Lu},
  {Hu}, {Li}, {Bai}, {Bian}, {Yuan}, {Ho}, {Wang}, \& {SEAMBH
  Collaboration}}]{Du18}
{Du}, P., {Zhang}, Z.-X., {Wang}, K., {et~al.} 2018, \apj, 856, 6

\bibitem[{{Gaskell} \& {Peterson}(1987)}]{Gaskell87}
{Gaskell}, C.~M., \& {Peterson}, B.~M. 1987, \apjs, 65, 1

\bibitem[{{Grier} {et~al.}(2013){Grier}, {Martini}, {Watson}, {Peterson},
  {Bentz}, {Dasyra}, {Dietrich}, {Ferrarese}, {Pogge}, \& {Zu}}]{Grier13}
{Grier}, C.~J., {Martini}, P., {Watson}, L.~C., {et~al.} 2013, \apj, 773, 90

\bibitem[{{Grier} {et~al.}(2017){Grier}, {Trump}, {Shen}, {Horne}, {Kinemuchi},
  {McGreer}, {Starkey}, {Brandt}, {Hall}, {Kochanek}, {Chen}, {Denney},
  {Greene}, {Ho}, {Homayouni}, {I-Hsiu Li}, {Pei}, {Peterson}, {Petitjean},
  {Schneider}, {Sun}, {AlSayyad}, {Bizyaev}, {Brinkmann}, {Brownstein},
  {Bundy}, {Dawson}, {Eftekharzadeh}, {Fernandez-Trincado}, {Gao},
  {Hutchinson}, {Jia}, {Jiang}, {Oravetz}, {Pan}, {Paris}, {Ponder}, {Peters},
  {Rogerson}, {Simmons}, {Smith}, \& {Wang}}]{Grier17}
{Grier}, C.~J., {Trump}, J.~R., {Shen}, Y., {et~al.} 2017, \apj, 851, 21

\bibitem[{Gunn {et~al.}(2006)Gunn, Siegmund, Mannery, Owen, Hull, Leger, Carey,
  Knapp, York, Boroski, Kent, Lupton, Rockosi, Evans, Waddell, Anderson, Annis,
  Barentine, Bartoszek, Bastian, Bracker, Brewington, Briegel, Brinkmann,
  Brown, Carr, Czarapata, Drennan, Dombeck, Federwitz, Gillespie, Gonzales,
  Hansen, Harvanek, Hayes, Jordan, Kinney, Klaene, Kleinman, Kron, Kresinski,
  Lee, Limmongkol, Lindenmeyer, Long, Loomis, McGehee, Mantsch, Eric
  H.~Neilsen, Neswold, Newman, Nitta, John~Peoples, Pier, Prieto, Prosapio,
  Rivetta, Schneider, Snedden, \& i~Wang}]{Gunn06}
Gunn, J.~E., Siegmund, W.~A., Mannery, E.~J., {et~al.} 2006, The Astronomical
  Journal, 131, 2332

\bibitem[{{Kaspi} {et~al.}(2005){Kaspi}, {Maoz}, {Netzer}, {Peterson},
  {Vestergaard}, \& {Jannuzi}}]{Kaspi05}
{Kaspi}, S., {Maoz}, D., {Netzer}, H., {et~al.} 2005, \apj, 629, 61

\bibitem[{{Kaspi} {et~al.}(2000){Kaspi}, {Smith}, {Netzer}, {Maoz}, {Jannuzi},
  \& {Giveon}}]{Kaspi00}
{Kaspi}, S., {Smith}, P.~S., {Netzer}, H., {et~al.} 2000, \apj, 533, 631

\bibitem[{{Kollmeier} {et~al.}(2017){Kollmeier}, {Zasowski}, {Rix}, {Johns},
  {Anderson}, {Drory}, {Johnson}, {Pogge}, {Bird}, {Blanc}, {Brownstein},
  {Crane}, {De Lee}, {Klaene}, {Kreckel}, {MacDonald}, {Merloni}, {Ness},
  {O'Brien}, {Sanchez-Gallego}, {Sayres}, {Shen}, {Thakar}, {Tkachenko},
  {Aerts}, {Blanton}, {Eisenstein}, {Holtzman}, {Maoz}, {Nandra}, {Rockosi},
  {Weinberg}, {Bovy}, {Casey}, {Chaname}, {Clerc}, {Conroy}, {Eracleous},
  {G{\"a}nsicke}, {Hekker}, {Horne}, {Kauffmann}, {McQuinn}, {Pellegrini},
  {Schinnerer}, {Schlafly}, {Schwope}, {Seibert}, {Teske}, \& {van
  Saders}}]{Kollmeier17}
{Kollmeier}, J.~A., {Zasowski}, G., {Rix}, H.-W., {et~al.} 2017, arXiv
  e-prints, arXiv:1711.03234

\bibitem[{{Kormendy} \& {Ho}(2013)}]{KormendyHo13}
{Kormendy}, J., \& {Ho}, L.~C. 2013, \araa, 51, 511

\bibitem[{{Li} {et~al.}(2019){Li}, {Shen}, {Brandt}, {Grier}, {Hall}, {Ho},
  {Homayouni}, {Horne}, {Schneider}, {Trump}, \& {Starkey}}]{Li19}
{Li}, I-Hsiu, J., {Shen}, Y., {Brandt}, W.~N., {et~al.} 2019, \apj, 884, 119

\bibitem[{{Onken} {et~al.}(2007){Onken}, {Valluri}, {Peterson}, {Pogge},
  {Bentz}, {Ferrarese}, {Vestergaard}, {Crenshaw}, {Sergeev}, {McHardy},
  {Merritt}, {Bower}, {Heckman}, \& {Wand el}}]{Onken07}
{Onken}, C.~A., {Valluri}, M., {Peterson}, B.~M., {et~al.} 2007, \apj, 670, 105

\bibitem[{{Pancoast} {et~al.}(2014){Pancoast}, {Brewer}, {Treu}, {Park},
  {Barth}, {Bentz}, \& {Woo}}]{Pancoast14}
{Pancoast}, A., {Brewer}, B.~J., {Treu}, T., {et~al.} 2014, \mnras, 445, 3073

\bibitem[{{Pei} {et~al.}(2017){Pei}, {Fausnaugh}, {Barth}, {Peterson}, {Bentz},
  {De Rosa}, {Denney}, {Goad}, {Kochanek}, {Korista}, {Kriss}, {Pogge},
  {Bennert}, {Brotherton}, {Clubb}, {Dalla Bont{\`a}}, {Filippenko}, {Greene},
  {Grier}, {Vestergaard}, {Zheng}, {Adams}, {Beatty}, {Bigley}, {Brown},
  {Brown}, {Canalizo}, {Comerford}, {Coker}, {Corsini}, {Croft}, {Croxall},
  {Deason}, {Eracleous}, {Fox}, {Gates}, {Henderson}, {Holmbeck}, {Holoien},
  {Jensen}, {Johnson}, {Kelly}, {Kim}, {King}, {Lau}, {Li}, {Lochhaas}, {Ma},
  {Manne-Nicholas}, {Mauerhan}, {Malkan}, {McGurk}, {Morelli}, {Mosquera},
  {Mudd}, {Muller Sanchez}, {Nguyen}, {Ochner}, {Ou-Yang}, {Pancoast}, {Penny},
  {Pizzella}, {Poleski}, {Runnoe}, {Scott}, {Schimoia}, {Shappee}, {Shivvers},
  {Simonian}, {Siviero}, {Somers}, {Stevens}, {Strauss}, {Tayar}, {Tejos},
  {Treu}, {Van Saders}, {Vican}, {Villanueva}, {Yuk}, {Zakamska}, {Zhu},
  {Anderson}, {Ar{\'e}valo}, {Bazhaw}, {Bisogni}, {Borman}, {Bottorff},
  {Brandt}, {Breeveld}, {Cackett}, {Carini}, {Crenshaw}, {De
  Lorenzo-C{\'a}ceres}, {Dietrich}, {Edelson}, {Efimova}, {Ely}, {Evans},
  {Ferland}, {Flatland}, {Gehrels}, {Geier}, {Gelbord}, {Grupe}, {Gupta},
  {Hall}, {Hicks}, {Horenstein}, {Horne}, {Hutchison}, {Im}, {Joner}, {Jones},
  {Kaastra}, {Kaspi}, {Kelly}, {Kennea}, {Kim}, {Kim}, {Klimanov}, {Lee},
  {Leonard}, {Lira}, {MacInnis}, {Mathur}, {McHardy}, {Montouri}, {Musso},
  {Nazarov}, {Netzer}, {Norris}, {Nousek}, {Okhmat}, {Papadakis}, {Parks},
  {Pott}, {Rafter}, {Rix}, {Saylor}, {Schn{\"u}lle}, {Sergeev}, {Siegel},
  {Skielboe}, {Spencer}, {Starkey}, {Sung}, {Teems}, {Turner}, {Uttley},
  {Villforth}, {Weiss}, {Woo}, {Yan}, {Young}, \& {Zu}}]{Pei17}
{Pei}, L., {Fausnaugh}, M.~M., {Barth}, A.~J., {et~al.} 2017, \apj, 837, 131

\bibitem[{{Peng} {et~al.}(2002){Peng}, {Ho}, {Impey}, \& {Rix}}]{Peng02}
{Peng}, C.~Y., {Ho}, L.~C., {Impey}, C.~D., \& {Rix}, H.-W. 2002, \aj, 124, 266

\bibitem[{{Peterson} {et~al.}(2004){Peterson}, {Ferrarese}, {Gilbert}, {Kaspi},
  {Malkan}, {Maoz}, {Merritt}, {Netzer}, {Onken}, {Pogge}, {Vestergaard}, \&
  {Wandel}}]{Peterson04}
{Peterson}, B.~M., {Ferrarese}, L., {Gilbert}, K.~M., {et~al.} 2004, \apj, 613,
  682

\bibitem[{{Richards} {et~al.}(2006){Richards}, {Lacy}, {Storrie-Lombardi},
  {Hall}, {Gallagher}, {Hines}, {Fan}, {Papovich}, {Vanden Berk}, {Trammell},
  {Schneider}, {Vestergaard}, {York}, {Jester}, {Anderson}, {Budav{\'a}ri}, \&
  {Szalay}}]{Richards06}
{Richards}, G.~T., {Lacy}, M., {Storrie-Lombardi}, L.~J., {et~al.} 2006, \apjs,
  166, 470

\bibitem[{{Ross} {et~al.}(2013){Ross}, {McGreer}, {White}, {Richards}, {Myers},
  {Palanque-Delabrouille}, {Strauss}, {Anderson}, {Shen}, {Brandt},
  {Y{\`e}che}, {Swanson}, {Aubourg}, {Bailey}, {Bizyaev}, {Bovy}, {Brewington},
  {Brinkmann}, {DeGraf}, {Di Matteo}, {Ebelke}, {Fan}, {Ge}, {Malanushenko},
  {Malanushenko}, {Mandelbaum}, {Maraston}, {Muna}, {Oravetz}, {Pan},
  {P{\^a}ris}, {Petitjean}, {Schawinski}, {Schlegel}, {Schneider}, {Silverman},
  {Simmons}, {Snedden}, {Streblyanska}, {Suzuki}, {Weinberg}, \&
  {York}}]{Ross13}
{Ross}, N.~P., {McGreer}, I.~D., {White}, M., {et~al.} 2013, \apj, 773, 14

\bibitem[{{Shen} \& {Ho}(2014)}]{Shen14}
{Shen}, Y., \& {Ho}, L.~C. 2014, \nat, 513, 210

\bibitem[{Shen {et~al.}(2011)Shen, Richards, Strauss, Hall, Schneider, Snedden,
  Bizyaev, Brewington, Malanushenko, Malanushenko, Oravetz, Pan, \&
  Simmons}]{Shen11}
Shen, Y., Richards, G.~T., Strauss, M.~A., {et~al.} 2011, The Astrophysical
  Journal Supplement Series, 194, 45

\bibitem[{{Shen} {et~al.}(2015{\natexlab{a}}){Shen}, {Greene}, {Ho}, {Brandt},
  {Denney}, {Horne}, {Jiang}, {Kochanek}, {McGreer}, {Merloni}, {Peterson},
  {Petitjean}, {Schneider}, {Schulze}, {Strauss}, {Tao}, {Trump}, {Pan}, \&
  {Bizyaev}}]{Shen15a}
{Shen}, Y., {Greene}, J.~E., {Ho}, L.~C., {et~al.} 2015{\natexlab{a}}, \apj,
  805, 96

\bibitem[{{Shen} {et~al.}(2015{\natexlab{b}}){Shen}, {Brandt}, {Dawson},
  {Hall}, {McGreer}, {Anderson}, {Chen}, {Denney}, {Eftekharzadeh}, {Fan},
  {Gao}, {Green}, {Greene}, {Ho}, {Horne}, {Jiang}, {Kelly}, {Kinemuchi},
  {Kochanek}, {P{\^a}ris}, {Peters}, {Peterson}, {Petitjean}, {Ponder},
  {Richards}, {Schneider}, {Seth}, {Smith}, {Strauss}, {Tao}, {Trump},
  {Wood-Vasey}, {Zu}, {Eisenstein}, {Pan}, {Bizyaev}, {Malanushenko},
  {Malanushenko}, \& {Oravetz}}]{Shen15b}
{Shen}, Y., {Brandt}, W.~N., {Dawson}, K.~S., {et~al.} 2015{\natexlab{b}},
  \apjs, 216, 4

\bibitem[{{Shen} {et~al.}(2019){Shen}, {Hall}, {Horne}, {Zhu}, {McGreer},
  {Simm}, {Trump}, {Kinemuchi}, {Brandt}, \& {Green}}]{Shen19}
{Shen}, Y., {Hall}, P.~B., {Horne}, K., {et~al.} 2019, \apjs, 241, 34

\bibitem[{Smee {et~al.}(2013)Smee, Gunn, Uomoto, Roe, Schlegel, Rockosi, Carr,
  Leger, Dawson, Olmstead, Brinkmann, Owen, Barkhouser, Honscheid, Harding,
  Long, Lupton, Loomis, Anderson, Annis, Bernardi, Bhardwaj, Bizyaev, Bolton,
  Brewington, Briggs, Burles, Burns, Castander, Connolly, Davenport, Ebelke,
  Epps, Feldman, Friedman, Frieman, Heckman, Hull, Knapp, Lawrence, Loveday,
  Mannery, Malanushenko, Malanushenko, Merrelli, Muna, Newman, Nichol, Oravetz,
  Pan, Pope, Ricketts, Shelden, Sandford, Siegmund, Simmons, Smith, Snedden,
  Schneider, SubbaRao, Tremonti, Waddell, \& York}]{Smee13}
Smee, S.~A., Gunn, J.~E., Uomoto, A., {et~al.} 2013, The Astronomical Journal,
  146, 32

\bibitem[{{Starkey} {et~al.}(2016){Starkey}, {Horne}, \&
  {Villforth}}]{Starkey16}
{Starkey}, D.~A., {Horne}, K., \& {Villforth}, C. 2016, \mnras, 456, 1960

\bibitem[{{Veilleux} \& {Osterbrock}(1987)}]{VO87}
{Veilleux}, S., \& {Osterbrock}, D.~E. 1987, \apjs, 63, 295

\bibitem[{{Wang} {et~al.}(2014{\natexlab{a}}){Wang}, {Qiu}, {Du}, \&
  {Ho}}]{Wang14c}
{Wang}, J.-M., {Qiu}, J., {Du}, P., \& {Ho}, L.~C. 2014{\natexlab{a}}, \apj,
  797, 65

\bibitem[{{Wang} {et~al.}(2014{\natexlab{b}}){Wang}, {Du}, {Hu}, {Netzer},
  {Bai}, {Lu}, {Kaspi}, {Qiu}, {Li}, {Wang}, \& {SEAMBH
  Collaboration}}]{Wang14b}
{Wang}, J.-M., {Du}, P., {Hu}, C., {et~al.} 2014{\natexlab{b}}, \apj, 793, 108

\bibitem[{{White} \& {Peterson}(1994)}]{White94}
{White}, R.~J., \& {Peterson}, B.~M. 1994, \pasp, 106, 879

\bibitem[{{Woo} {et~al.}(2015){Woo}, {Yoon}, {Park}, {Park}, \& {Kim}}]{Woo15}
{Woo}, J.-H., {Yoon}, Y., {Park}, S., {Park}, D., \& {Kim}, S.~C. 2015, \apj,
  801, 38

\bibitem[{{Yip} {et~al.}(2004){Yip}, {Connolly}, {Vanden Berk}, {Ma},
  {Frieman}, {SubbaRao}, {Szalay}, {Richards}, {Hall}, {Schneider}, {Hopkins},
  {Trump}, \& {Brinkmann}}]{Yip04}
{Yip}, C.~W., {Connolly}, A.~J., {Vanden Berk}, D.~E., {et~al.} 2004, \aj, 128,
  2603

\bibitem[{{Yu} {et~al.}(2019){Yu}, {Bian}, {Wang}, {Zhao}, \& {Ge}}]{Yu19}
{Yu}, L.-M., {Bian}, W.-H., {Wang}, C., {Zhao}, B.-X., \& {Ge}, X. 2019,
  \mnras, 488, 1519

\bibitem[{{Yu} {et~al.}(2020){Yu}, {Kochanek}, {Peterson}, {Zu}, {Brandt},
  {Cackett}, {Fausnaugh}, \& {McHardy}}]{Yu20}
{Yu}, Z., {Kochanek}, C.~S., {Peterson}, B.~M., {et~al.} 2020, \mnras, 491,
  6045

\bibitem[{{Yue} {et~al.}(2018){Yue}, {Jiang}, {Shen}, {Hall}, {Yu},
  {Schneider}, {Ho}, {Horne}, {Petitjean}, \& {Trump}}]{Yue18}
{Yue}, M., {Jiang}, L., {Shen}, Y., {et~al.} 2018, \apj, 863, 21

\bibitem[{{Zu} {et~al.}(2011){Zu}, {Kochanek}, \& {Peterson}}]{Zu11}
{Zu}, Y., {Kochanek}, C.~S., \& {Peterson}, B.~M. 2011, \apj, 735, 80

\end{thebibliography}

\end{document}